\documentclass[aps,prd,preprint,showpacs,eqsecnum,nofootinbib]{revtex4-2}
\usepackage{graphicx,epsf,subfigure,epstopdf}
\usepackage{amsmath,xfrac,amssymb}
\usepackage{amsfonts}
\usepackage{MnSymbol} 
\usepackage{scalerel}
\usepackage{cancel}
\usepackage{mathtools}
\usepackage{accents,placeins}
\usepackage[usenames,dvipsnames]{xcolor}
\usepackage{slashed}
\usepackage{cancel}

\usepackage[export]{adjustbox}
\usepackage{titlesec}
\usepackage{setspace} 
\usepackage{xpatch}
\usepackage[obeyDraft]{todonotes}

\usepackage[final]{hyperref}
\hypersetup{
    colorlinks=true,
    linkcolor=blue,
    citecolor=Green,
    urlcolor=blue,
    linktoc=page
}   

\newif\ifdraft
\drafttrue
\newif\ifpreprint
\preprinttrue

\def\fig#1{fig.~{\ref{#1}}}

\def\figs#1#2{figs.~{\ref{#1}} and {\ref{#2}}}

\def\Sect#1{Section~{\ref{#1}}}
\def\sect#1{section~{\ref{#1}}}
\def\sects#1#2{sections~{\ref{#1}} and~{\ref{#2}}}
\def\sectst#1#2#3{sections~{\ref{#1}},~{\ref{#2}}, and~{\ref{#3}}}

\def\Tab#1{Table~{\ref{#1}}}

\def\spa#1.#2{\left\langle#1\,#2\right\rangle}
\def\spb#1.#2{\left[#1\,#2\right]}
\def\spash#1.#2{\spa{\smash{#1}}.{\smash{#2}}}
\def\spbsh#1.#2{\spb{\smash{#1}}.{\smash{#2}}}
\def\lor#1.#2{\left(#1\,#2\right)}
\def\sand#1.#2.#3{%
\left\langle\smash{#1^{-}}{\vphantom1}\right|{#2}%
\left|\smash{#3^{-}}{\vphantom1}\right\rangle}
\def\sandpp#1.#2.#3{%
\left\langle\smash{#1^{+}}{\vphantom1}\right|{#2}%
\left|\smash{#3^{+}}{\vphantom1}\right\rangle}
\def\sandpm#1.#2.#3{%
\left\langle\smash{#1^{+}}\vphantom1\right|{#2}%
\left|\smash{#3^{-}}\vphantom1\right\rangle}
\def\sandmp#1.#2.#3{%
\left\langle\smash{#1^{-}}\vphantom1\right|{#2}%
\left|\smash{#3^{+}}{\vphantom1}\right\rangle}
\def\sandmppm#1.#2.#3{%
\left\langle\smash{#1^{\mp}}\vphantom1\right|{#2}%
\l
eft|\smash{#3^{\pm}}{\vphantom1}\right\rangle}
\def\sandnn#1.#2.#3{%
\left\langle\smash{#1}\vphantom1\right|{#2}%
\left|\smash{#3}{\vphantom1}\right\rangle}
\def\sandmn#1.#2.#3{%
\left\langle\smash{#1^{-}}\vphantom1\right|{#2}%
\left|\smash{#3}{\vphantom1}\right\rangle}
\def\sandnm#1.#2.#3{%
\left\langle\smash{#1}\vphantom1\right|{#2}%
\left|\smash{#3^{-}}{\vphantom1}\right\rangle}

\def\nsand#1.#2.#3{%
        \left\langle\smash{#1}{\vphantom1}\right|{#2}%
        \left|\smash{#3}{\vphantom1}\right]}
\def\nsandaa#1.#2.#3{%
	\left\langle\smash{#1}{\vphantom1}\right|{#2}%
	\left|\smash{#3}{\vphantom1}\right\rangle}
\def\nsandbb#1.#2.#3{%
	\left[\smash{#1}{\vphantom1}\right|{#2}%
	\left|\smash{#3}{\vphantom1}\right]}
\def\nsandba#1.#2.#3{%
	\left[\smash{#1}{\vphantom1}\right|{#2}%
	\left|\smash{#3}{\vphantom1}\right\rangle}

\def\Tr{\, {\rm Tr}}

\def\eps{\epsilon}

\def\tab#1{table~\ref{#1}}

\def\eqn#1{eq~(\ref{#1})}
\def\Eqn#1{Equation~(\ref{#1})}
\def\eqns#1#2{eqs.~(\ref{#1}) and~(\ref{#2})}

\def\be{\begin{equation}}
\def\ee{\end{equation}}

\def\onehalf{\frac12}

\def\Ord{{\cal O}}

\newbox\charbox
\newbox\slabox
\def\s#1{{      
        \setbox\charbox=\hbox{$#1$}
        \setbox\slabox=\hbox{$/$}
        \dimen\charbox=\ht\slabox
        \advance\dimen\charbox by -\dp\slabox
        \advance\dimen\charbox by -\ht\charbox
        \advance\dimen\charbox by \dp\charbox
        \divide\dimen\charbox by 2
        \raise-\dimen\charbox\hbox to \wd\charbox{\hss/\hss}
        \llap{$#1$} }}

\def\simasO^#1{\stackrel{#1}{\sim}}
\def\simas^#1{\hskip -3pt%
\stackrel{#1}{\lower 5pt\hbox{$\widetilde{\hphantom{#1}}$}}%
\hskip -3pt}

\newcommand\gatop[2]{\genfrac{}{}{0pt}{}{#1}{#2}}

\def\Singular{\textsf{Singular\/}}

\begin{document}

\def\vec#1{\boldsymbol{#1}}
\def\LIPS{\mathop{\rm LIPS}\nolimits}
\def\LIPSS{\LIPS_s}
\def\matelt#1#2{\mathcal{M}^{(#1)}_{#2}}
\def\nlomat#1{\mathcal{\widehat M}^{\rm NLO}_{#1}}
\def\netmatelt#1#2{\mathcal{\widehat M}^{(#1)}_{#2}}
\def\virtual{{\rm V}}
\def\real{{\rm R}}
\def\differ{{\rm \Delta}}
\def\modreal{{\rm\hat R}}
\def\realT{{\modreal}}
\def\nlo{{\rm NLO}}
\def\nloT{\widehat{\rm NLO}}
\def\djets#1#2{d^{#1} J_1\cdots d^{#1} J_{#2}}
\def\dvecjets#1#2{d^{#1} \vec{J}_1\cdots d^{#1} \vec{J}_{#2}}

\def\triskelion#1#2#3{(#1\,|\,#2,#3)}

\hfuzz=15 pt
\begin{minipage}{\textwidth}

\title{Universal Decomposition of Phase-Space Integrands}

\def\squeezev{\vspace*{-2.5mm}}

\author{David~A.~Kosower}
\affiliation{\looseness=-1%
\linespread{1}\selectfont%
Theoretical Physics Department, CERN, CH--1211 Geneva~23, Switzerland
\\ \textrm{\rm and}\\Institut de Physique Th\'eorique, CEA, CNRS, Universit\'e Paris--Saclay,
  F--91191 Gif-sur-Yvette cedex, France
  \\ {\sf David.Kosower@ipht.fr}
}

\author{Ben Page}
\affiliation{\looseness=-1%
\linespread{1}\selectfont%
Theoretical Physics Department, CERN, CH--1211 Geneva~23, Switzerland
\\ {\sf Ben.Page@cern.ch}
}
\date\today

\begin{abstract}
One-loop integrands can be written in terms of a simple,
process-independent basis.  We show that a similar basis 
exists for integrands of phase-space integrals for the real-emission
contribution at next-to-leading order.  Our demonstration 
deploys techniques from computational algebraic geometry to
partial-fraction integrands in a systematic way.
This takes the first step towards a decomposition of phase-space integrals
in terms of a basis of master integrals.
\end{abstract}

\pacs{\hspace{1cm}}

\maketitle
\end{minipage}

\newpage
\thispagestyle{empty}
\tableofcontents
\newpage

\section{Introduction}
\label{IntroSection}

Theoretical calculations of Standard-Model processes have played an 
important role in
extracting information from experiments at both electron--positron and 
hadron colliders.
This is especially true for calculations in perturbative QCD.  These 
calculations
are needed to measure Standard-Model parameters from data, and to 
confront data with
expectations in the search for new physics.  Nature has not been so 
kind to date as to reveal
direct evidence of physics beyond the Standard Model at CERN's Large 
Hadron Collider (LHC), in
spite of the tantalizing hints from the existence of dark matter and 
the plausibility of
leptogenetic origins to the observed astrophysical
baryon--antibaryon asymmetry.

Experiments are thus proceeding into an era of precision 
measurements, driven by the
prospects of increasing luminosity at the LHC and the increasing 
success of experiments
at overcoming issues arising from multi-event 
pileup~\cite{Pileup}.  This path calls 
for increasing
precision in theoretical predictions, in order to minimize the part of 
theoretical uncertainties
in the error budget of measurements and resulting constraints.   
Within a perturbative expansion,
the large and quickly running QCD coupling $\alpha_s$ means that 
efficiently predictions, especially for multi-jet observables, are only 
qualitative; that next-to-leading
order (NLO) predictions are the first quantitative order; and that 
precision predictions
require next-to-next-to-leading order (NNLO) predictions.

NNLO predictions require computations of two-loop amplitudes, a 
subject which has seen a great deal of progress in recent years. 
This progress has come on two fronts. Firstly, our understanding 
of how to compute the master integrals efficiently 
(\textit{e.g.\/}~with pentagon 
functions~\cite{Gehrmann:2018yef, Chicherin:2020oor,Chicherin:2021dyp})
within the differential equations 
method~\cite{Kotikov:1990kg, Remiddi:1997ny, Henn:2013pwa} 
has improved. 
Secondly, finite-field 
methods~\cite{vonManteuffel:2014ixa,Peraro:2016wsq} 
(see also recent work on extension to $p$-adic 
numbers~\cite{DeLaurentis:2022otd}) have made 
it practical to obtain analytic forms for amplitudes using numerical 
approaches.   In particular,
this has facilitated the application of on-shell approaches, 
such as numerical 
unitarity~\cite{Abreu:2017hqn, Abreu:2017xsl, Abreu:2020xvt}, 
which build on earlier dramatic progress in the computation of one-loop 
amplitudes~\cite{NLOProgress}.
With these technical advances, the community has been able to  
compute numerous two-loop five-point amplitudes (see 
refs.~\cite{Abreu:2021asb, Badger:2022ncb} 
for recent cutting-edge five-point-one-mass examples).

NNLO predictions also require the computation of real-emission 
integrals, and
combining them with the two-loop virtual contributions.  Each of these 
contributions
is singular.  The singularities are in practice always regulated 
dimensionally.
They are manifest in the virtual contributions.
In contrast, they arise entirely from
the integration over phase-space in the double-emission contributions 
and partly from such integrations in the single-emission
ones.  A direct computation of them appears intractable, and so 
various strategies
have been proposed to isolate the singularities into simpler 
integrals which are
tractable analytically.  These strategies fall into three main 
groups: slicing, subtraction, and hybrids of the two.

In a slicing approach~\cite{Slicing}, 
the integrals over the singular (soft and collinear)
regions of phase space are performed analytically using
the singular approximations to the matrix element.  These integrals
are universal because the soft and collinear approximations
are universal.  
The integral
over the remainder of phase space is done numerically.  The 
definition of the singular regions can be done using pairs of
invariants at the level of color-ordered amplitudes, or using
other measures such as $N$-jettiness~\cite{NJettiness}.  
The slicing approach saw early use in NLO
predictions of jet production.
In a subtraction 
approach~\cite{CataniSeymour,FKS,NagySoperSubtraction}, 
one uses the universal functions describing the soft and collinear 
behavior of the matrix elements to build subtraction terms rendering 
the integrand finite everywhere in phase space.  The subtracted 
integral is done numerically, while
the integrals of the subtraction terms are performed analytically
over all of phase space.  The approach of ref.~\cite{CataniSeymour}
in particular became the most widely used one at NLO.

A hybrid approach has been used extensively at NNLO, where a
subtraction effectively controls the NLO matrix element, and
$N$-jettiness slicing or 
$q_\textrm{T}$-slicing~\cite{qTNNLOSubtraction}
controls its use in the singular regions.
There are also NNLO versions of subtraction approaches,
including antenna subtraction~\cite{AntennaSubtractionNNLO}, 
and approaches~\cite{ColorfulSubtraction,SectorResidueSubtraction,%
LocalAnalyticSubtraction}
based on generalizations of NLO strategies.

Deployment of these strategies at NNLO has proven far more 
difficult than might have
been hoped, given the experience gleaned from their successful 
use and automation
at NLO.  This has motivated us to search for a new approach, 
which seeks to tackle 
the seemingly intractable problem of computing phase-space 
integrals directly, allowing
for numerical computation of cross section while maintaining 
analytic control of
singular terms.  We are also motivated by the goal of implementing 
an approach that
will allow
computing a fully
differential cross section at NLO, echoing the success of the 
`matrix-element method at NLO'
approach of refs.~\cite{MEM@NLO}.
In this paper, we take the first steps towards this goal.  
We explore an approach based on decomposing phase-space integrands into
simpler integrals, analogs of master integrands in loop computations.  
In the present paper, we will consider the problem at NLO.  

At this order, we are interested in integrating over the emission of 
a lone gluon
(or the splitting of a gluon into a quark--antiquark pair). 
Ultimately, we would seek to show that
phase-space integrands from arbitrary processes at NLO can be 
reduced to a limited set,
with no integrand term containing more than four denominator
factors dependent 
on this lone
gluon, and hence contributing to infrared
or collinear singularities. In addition, we would seek
to show reducibility of numerators.  These simplifications 
are analogous to the well-known
reductions of one-loop integrals to a superset of bubble, triangle, 
and box integrands.

\def\lamh{{\hat\lambda}}

The phase-space integrands are matrix elements, built at
lowest order from squares of tree amplitudes.
In this paper, we will take the very first step in this direction by
showing that tree amplitudes of arbitrary multiplicity can be decomposed
into sums of rational functions where each term contains no more
than four denominators dependent on the singular gluon.  This
shows that there is a process-independent
finite basis of
terms in both the tree amplitude and in the squared matrix
element.  We leave the question of reducing this basis to
a minimal one to future work.

In the case of loop integrals, simple algebra and a consideration 
of appropriately
constructed numerators suffice to demonstrate this reduction.  
The reduction for phase-space
integrands is more difficult.  We employ a masslessness-preserving
remapping of partonic momenta along with
techniques from 
computational algebraic geometry to carry it out.

In the next section, we present the setting for our investigation,
that of jet observables in heavy-particle decay.  The reader will
find a more detailed roadmap to the remainder of the article at its 
end.  The road starts 
in \sect{AntennaFactorizationSection}, where we review antenna 
factorization.  In \sect{DimensionalRegularizationSection},
we discuss some subtleties concerning dimensional regularization
and counting of degrees of freedom.  
In \sect{InverseAntennaMappingSection}, we invert the antenna mapping.
We use the inverse to building required
changes of variables for later sections.  
In 
\sect{OneLoopDecompositionSection}, we review
general aspects of reducing one-loop integrands.  
We review the standard reduction of integrands with constant
numerators in \sect{LoopNumeratorFreeReductionSection}
and recast this reduction
in the language of algebraic geometry.  We present
a brief review of computational algebraic geometry in 
\sect{ComputationalAlgebraicGeometrySection}.
We give two examples of reducing contributions to
tree-level amplitudes 
in \sect{RealEmissionNumeratorFreeReductionSection}.  
In \sect{TriskeliaSection},
we introduce the notion of \textsl{triskelia\/} to classify all
different contributions to an amplitude.
We survey the reduction of all such different contributions
in \sect{ResultsSection}.  In \sect{NumeratorReductionSection}, 
we consider the reduction of
integrand numerators, recasting the loop-integral case
in terms of computational algebraic geometry, and applying
that recasting to tree-level amplitudes.  We make a few
concluding remarks in \sect{ConclusionSection}.

\section{Jet Observables}
\label{JetObservablesSection}

\def\Obs{\mathcal{O}}
\def\obs{v}
\def\JetRecomb{{\mathop{\rm Rec}\nolimits}}
\def\BasicJetTheta{{\mathop{\rm Rec}\nolimits}^J}
\def\DurhamJetRecomb{\JetRecomb^{\textrm{Dur}}}
\def\CutsTheta{C^J}

In collider applications of perturbative QCD, differential
cross sections of various observables $\Obs$ built out of jets 
play an important role.
Their prediction depends on a jet algorithm and on phase-space
cuts on the jets and and other final-state objects.  In this section, 
we use the decay of a colorless scalar $\phi$ 
as an example to set up the 
framework for our investigation into simplifying the 
real-emission contributions.  
We take it to couple to
gluons via a $f_0 \phi G^2$, where $G_{\mu\nu}$ is the gluon field
strength and the coupling $f_0$ has dimensions of
inverse energy.

\def\JSum{J_\Sigma}
\def\deltap{\delta^{(+)}}
The singly differential cross section in a jet observable 
$\Obs$ built out of $n$ jet momenta $J_i$ has the form,
\begin{equation}
\frac{d\sigma}{d\obs} = 
\int d\LIPS^D_n\bigl(K_0;\{J_i\}_{i=1}^n\bigr)\,
\,\delta\bigl(\obs-\Obs(\{J_i\}_{i=1}^n)\bigr)\,
\CutsTheta_n\bigl(\{J_i\}_{i=1}^n\bigr)
\frac{d^{nD}\sigma}{\djets{D}{n}}\,.
\label{DifferentialCrossSection}
\end{equation}
The decaying scalar has four-momentum $K_0$, and the 
$D$-dimensional Lorentz-invariant phase-space measure is,
\begin{equation}
d\LIPS^D_n\bigl(K_0;\{J_i\}_{i=1}^n\bigr) \equiv 
\prod_{i=1}^n \frac{d^D J_i}{(2\pi)^3} \deltap(J_i^2)
\,\delta^{(D)}\Bigl(K_0-\sum_{i=1}^n J_i\Bigr)\,.
\label{LorentzInvariantPhaseSpaceMeasure}
\end{equation}
The jets are taken to be massless, as is typical of a modern jet 
algorithm using a massless recombination scheme such as 
the $E_\textrm{T}$
scheme.
We take the jet algorithm to be infrared- and collinear-safe (IRCS).
The function $\CutsTheta_n$ imposes cuts on the protojets.
The fully differential cross section with respect to all
jets is ${d^{nD}\sigma}/{\djets{D}{n}}$.
It can be expanded order by order in
perturbation theory,
\begin{equation}
\frac{d^{nD}\sigma}{\djets{D}{n}} =
\frac{d^{nD}\sigma^{\rm LO}}{\djets{D}{n}} + 
\frac{d^{nD}\sigma^{\rm NLO}}{\djets{D}{n}} +
\frac{d^{nD}\sigma^{\rm NNLO}}{\djets{D}{n}} +
\cdots\,,    
\end{equation}
where the leading-order (LO) term is $\Ord(f_0\alpha^{n-2})$,
the next-to-leading order (NLO) term is $\Ord(f_0\alpha^{n-1})$,
the next-to-next-leading order (NNLO) term 
is $\Ord(f_0\alpha^{n})$, and so on.

The most common type of jet algorithm is hierarchical.
We may distinguish three phases in such algorithms: 
selecting the partons or protojets
to be combined into a new level of protojets; combining the 
four-momenta of those protojets
into the new protojet four-momentum; and after reaching a stopping 
criterion, imposing
cuts on the protojets to yield the set of final jet four-momenta.
The selection procedure is typically implemented via a product of
theta functions, while the
combination procedure is implemented via a product of delta
functions.  (We retain
the traditional name of `recombination' for the latter.)  
In \eqn{DifferentialCrossSection}, it
is the $\CutsTheta_n$ function that
imposes the required cuts on protojet energies and rapidities, while 
the two first phases are
implicit in fully differential cross section $d^{nD}\sigma$.

At LO, the fully differential cross section is,
\begin{equation}
\begin{aligned}
\frac{d^{nD}\sigma^{\rm LO}}{\djets{D}{n}} &= 
\int d\LIPS^D_n\bigl(\JSum;\{k_i\}_{i=1}^n\bigr)\,
\JetRecomb\bigl(\{J_i\}_{i=1}^n\leftarrow\{k_i\}_{i=1}^n\bigr)\,
\\ & \hphantom{= \int}\hspace*{15mm}\times
\matelt{0}{n}\bigl(\JSum\rightarrow\{k_i\}_{i=1}^n\bigr)
\,.
\end{aligned}
\label{LOFullyDifferentialCrossSection}
\end{equation}
where $\JSum=\Sigma_i^n J_i$ (formally independent of $K_0$).
In this equation, 
$\matelt{0}{n}\bigl(\JSum\rightarrow\{k_i\}_{i=1}^n\bigr)$
is the squared tree-level matrix element for the decay of 
the colorless scalar carrying $K_0$ into
$n$ partons carrying the given momenta $k_i$.
The jet function 
$\JetRecomb$ represents the selection and 
recombination phases of the 
idealized jet algorithm,
with weight localized on configurations yielding exactly $n$ jets
with jet four-momenta $\{J_i\}_{i=1}^n$,
and 0 elsewhere.  
At LO, the protojets are simply the original partons,
and the jet recombination algorithm is given 
by a product of
delta functions $\delta^{(D)}(J_i-k_i)$.  The integrals are 
then trivial, and the differential
cross section is the same as the squared tree-level matrix element,
\begin{equation}
\frac{d^{nD}\sigma^{\rm LO}}{\djets{D}{n}} = 
\matelt{0}{n}\bigl(\JSum\rightarrow\{J_i\}_{i=1}^n\bigr)
\,.
\label{LOFullyDifferentialCrossSectionFinal}
\end{equation}
Beyond LO, the jet 
algorithm will of course contain nontrivial recombinations of partonic 
momenta.  

As an example, the $k_\perp$ algorithm (\textit{aka} the Durham
algorithm~\cite{DurhamAlgorithm}) in the three-parton
case would use a distance measure,
\begin{equation}
    \Delta_{ij} \equiv \min(k_i^0,k_j^0)^2 (1-\cos\theta_{ij})\,,
\end{equation}
with the recombination taking the following form at NLO,
\begin{equation}
    \begin{aligned}
    \DurhamJetRecomb&=\delta^{(D)}(J_2-k_3)\,\Theta(y-\Delta_{12})
    \Theta(\Delta_{23}-\Delta_{12})\Theta(\Delta_{13}-\Delta_{12})
    \\&\hspace*{11mm}\times\delta^{(D-1)}(\vec J_1-\vec k_1-\vec k_2)
    \delta(J_1^0 - |\vec k_1+\vec k_2|)
    \\& + (\textrm{permutations\ of\ }k_1, k_2, k_3)\,.
    \end{aligned}
\end{equation}
Here, $y$ is the jet-size parameter, and bold symbols denote 
spatial vectors.

The observable in \eqn{DifferentialCrossSection} is infrared- and 
collinear-safe by 
virtue of being a function of jet momenta resulting from an IRCS jet 
algorithm.
More general IRCS observables, such as the distribution of
jet energy radii, are certainly possible.
The decompositions we will develop
in this paper would apply to them as well.

The expressions in 
\eqns{DifferentialCrossSection}{LOFullyDifferentialCrossSection}
contain objects, such as the fully differential jet cross
section ${d^{nD}\sigma}/{\djets{D}{n}}$ and the recombination
function 
$\JetRecomb\bigl(\{J_i\}_{i=1}^n\leftarrow\{k_i\}_{i=1}^n\bigr)$,
which are themselves delta distributions.
In practical applications, each must be integrated over
bins corresponding to those used in an experimental analysis.

One of our goals is to give a practical and analytically tractable 
definition of
the fully differential cross section beyond leading order.  This 
is similar
to the goal in developments of the matrix-element method at 
NLO~\cite{MEM@NLO}.
Our first
task is to recast differential distributions in the form of 
\eqn{DifferentialCrossSection}.
Let us first recall the conventional approach to higher-order 
corrections to 
distributions. 
At next-to-leading order (NLO), for example, we will have two 
contributions, virtual
and real-emission.
We can write the virtual contribution in a form paralleling 
the all-orders 
form~(\ref{DifferentialCrossSection}).
 The virtual contribution is,
\begin{equation}
\frac{d\sigma^{\rlap{$\scriptstyle\virtual$}{}}}{d\obs} = 
\int d\LIPS^D_n\bigl(K_0;\{J_i\}_{i=1}^n\bigr)\,
\,\delta\bigl(\obs-\Obs(\{J_i\}_{i=1}^n)\bigr)\,
\CutsTheta_n\bigl(\{J_i\}_{i=1}^n\bigr)
\frac{d^{nD}\sigma^{\virtual}}{\djets{D}{n}}\,,
\label{NLOVirtualDifferentialCrossSection}
\end{equation}
where,
\begin{equation}
\begin{aligned}
\frac{d^{nD}\sigma^{\virtual}}{\djets{D}{n}} &= 
\int d\LIPS^D_n\bigl(\JSum;\{k_i\}_{i=1}^n\bigr)\,
\JetRecomb\bigl(\{J_i\}_{i=1}^n\leftarrow\{k_i\}_{i=1}^n\bigr)\,
\\ & \hphantom{= \int}\hspace*{15mm}\times
\matelt{1\virtual}{n}\bigl(\JSum\rightarrow\{k_i\}_{i=1}^n\bigr)
\\ &= 
\matelt{1\virtual}{n}\bigl(\JSum\rightarrow\{J_i\}_{i=1}^n\bigr)
\,.
\end{aligned}
\label{NLOVirtualFullyDifferentialCrossSection}
\end{equation}
Here, $\matelt{1\virtual}{n}$ is the interference of one-loop and 
tree-level matrix elements
for the decay of the colorless scalar into $n$ partons.
The real-emission contribution is similarly given by,
\begin{equation}
\frac{d\sigma^{\rlap{$\scriptstyle\real$}{}}}{d\obs} = 
\int d\LIPS^D_n\bigl(K_0;\{J_i\}_{i=1}^n\bigr)\,
\,\delta\bigl(\obs-\Obs(\{J_i\}_{i=1}^n)\bigr)\,
\CutsTheta_n\bigl(\{J_i\}_{i=1}^n\bigr)
\frac{d^{nD}\sigma^{\real}}{\djets{D}{n}}\,,
\label{NLORealDifferentialCrossSection}
\end{equation}
where
\begin{equation}
\begin{aligned}
\frac{d^{nD}\sigma^{\real}}{\djets{D}{n}}\, &= 
\int d\LIPS^D_{n+1}\bigl(\JSum;\{k_i\}_{i=1}^{n+1}\bigr)\,
\JetRecomb\bigl(\{J_i\}_{i=1}^n\leftarrow\{k_i\}_{i=1}^{n+1}\bigr)\,
\\ & \hphantom{= \int}\hspace*{15mm}\times
\matelt{0}{n+1}\bigl(\JSum\rightarrow\{k_i\}_{i=1}^{n+1}\bigr)\,.
\end{aligned}
\label{NLORealFullyDifferentialCrossSection}
\end{equation}
This contribution again depends solely on tree-level matrix elements, 
but this time
for the decay of the colorless scalar into $n+1$ partons.
In general, both the virtual and real-emission contributions
have $\eps^{-2}$ singularities ($D=4-2\eps$ as usual), which will 
cancel in
any physical observable thanks to the KLN theorem~\cite{KLN}.  
If we are somehow able to calculate the integrals in the fully
differential cross sections, then we could cancel singularities point 
by point
in the jet phase space, which is one of our goals.  

\def\Jac{\mathop{\rm Jac}\nolimits}
\def\pjet{\hat k}
The integrals in 
\eqns{NLOVirtualFullyDifferentialCrossSection}%
{NLORealFullyDifferentialCrossSection}
are taken over different partonic phase spaces, which
makes cancellation of singularities difficult.  We can, however, 
recast \eqn{NLORealFullyDifferentialCrossSection} in a form that 
makes it easier to discuss cancellations.
We begin by changing variables
in the real integration to one over $n$ protojets $\pjet_i$ and 
one `singular'
momentum $k_r$,
\begin{equation}
\begin{aligned}
\pjet_i &= \pjet_i\bigl(\{k_j\}_{j=1}^{n+1}\bigr)\,,
\\k_r &= k_r\bigl(\{k_j\}_{j=1}^{n+1}\bigr)\,.
\end{aligned}    
\end{equation}
The protojet and singular momenta are all massless.
The decomposition of the phase-space measure is exact,
\begin{equation}
\begin{aligned}
d\LIPS^D_{n+1}&\bigl(K;\{k_i\}_{i=1}^{n+1}\bigr) = 
\\ &
d\LIPS^D_n\bigl(K;\{\pjet_i\}_{i=1}^{n}\bigr)\,d\LIPSS^D(k_r)\,
\Jac\bigl(\{\pjet_i\}_{i=1}^n,k_r\bigr)
\\ &= \prod_{j=1}^n \frac{d^D \pjet_j}{(2\pi)^3} \deltap(\pjet_j^2)
\,\delta^{(D)}\Bigl(K-\sum_{j=1}^n \pjet_j\Bigr)
\,\frac{d^D k_r}{(2\pi)^3} 
\deltap(k_r^2)\,\Jac\bigl(\{\pjet_i\}_{i=1}^n,k_r\bigr)
\end{aligned}
\label{PhaseSpaceDecomposition}
\end{equation}
Here, $d\LIPSS$ indicates the singular phase space, and
$\Jac$ is the Jacobian from the change of variables.  The limits of 
integration
on $k_r$ are determined by the original phase-space boundaries.
The idea is that (within a slice of phase space)
all singularities of an integral are associated to 
specific regions of the $k_r$ integration, without reference
to integrations over the remaining $\pjet_i$.
We will introduce a concrete form for a suitable remapping below.
We can think of the mapping from the original momenta to
the protojets as an alternate, theory-friendly jet algorithm.

The change of variables is (in principle) invertible\footnote{This is 
true at NLO even for `standard' jet algorithms like $k_{\textrm{T}}$ or 
anti-$k_{\textrm{T}}$~\cite{AntiKT} in a
decay process.}, so that we can think of the
original partonic momenta as functions of the protojet momenta $\pjet_i$ 
along with $k_r$,
\begin{equation}
k_j = k_j\bigl(\{\pjet_i\}_{i=1}^n,k_r\bigr)\,.
\label{InverseChangeOfVariables}
\end{equation}
The protojet momenta $\pjet_i$ should be thought of as variables
on a par with the partonic momenta at LO: massless jet momenta before 
the event is passed through jet cuts.

Introducing the protojet momenta, we rewrite the radiative
contribution,
\begin{equation}
\begin{aligned}
\frac{d^{nD}\sigma^{\real}}{\djets{D}{n}}\, &= 
\int d\LIPS^D_{n}\bigl(\JSum;\{\pjet_i\}_{i=1}^{n}\bigr)\,
\,\int d\LIPSS^D(k_r)\,
\Jac\bigl(\{\pjet_i\}_{i=1}^n,k_r\bigr)\,
\\&\hphantom{= \int}\hspace*{5mm}\times
\JetRecomb\bigl(\{J_i\}_{i=1}^n\leftarrow
    \{k_i(\{\pjet_i\}_{i=1}^n,k_r)\}_{i=1}^{n+1}\bigr)\,
\matelt{0}{n+1}\bigl(\JSum\rightarrow\{k_i\}_{i=1}^{n+1}\bigr)\,.
\end{aligned}
\label{NLORealFullyDifferentialCrossSection2}
\end{equation}

\def\JetRecombT{\JetRecomb^{\textrm{T}}}
The jet algorithm here is the theoretical idealization of
the experimental jet algorithm.  We can now subtract and add
back a function implementing a different jet algorithm,
$\JetRecombT$, to obtain,
\begin{equation}
\frac{d^{nD}\sigma^{\real}}{\djets{D}{n}} =    
\frac{d^{nD}\sigma^{\differ}}{\djets{D}{n}}
+\frac{d^{nD}\sigma^{\realT}}{\djets{D}{n}}\,,
\end{equation}
where,
\begin{equation}
\begin{aligned}
\frac{d^{nD}\sigma^{\differ}}{\djets{D}{n}}\, &=
\int d\LIPS^D_{n}\bigl(\JSum;\{\pjet_i\}_{i=1}^{n}\bigr)\,
\,\int d\LIPSS^D(k_r)\,
\Jac\bigl(\{\pjet_i\}_{i=1}^n,k_r\bigr)\,
\\&\hphantom{=\int}\hspace*{5mm}\times
\bigl[\JetRecomb\bigl(\{J_i\}_{i=1}^n\leftarrow
    \{k_i(\{\pjet_i\}_{i=1}^n,k_r)\}_{i=1}^{n+1}\bigr)
\\&\hphantom{=\int}\hspace*{5mm}\hphantom{\times\bigl[]}
    -\JetRecombT\bigl(\{J_i\}_{i=1}^n\leftarrow
    \{k_i(\{\pjet_i\}_{i=1}^n,k_r)\}_{i=1}^{n+1}\bigr)\bigl]
    \,
\\ &\hphantom{=\int}\hspace*{5mm}\times
\matelt{0}{n+1}\bigl(\JSum\rightarrow
  \{k_j(\{\pjet_i\}_{i=1}^n,k_r)\}_{j=1}^{n+1}\bigr)
\,,
\\ \frac{d^{nD}\sigma^{\realT}}{\djets{D}{n}}\,&=
\int d\LIPS^D_{n}\bigl(\JSum;\{\pjet_i\}_{i=1}^{n}\bigr)\,
\,\int d\LIPSS^D(k_r)\,
\Jac\bigl(\{\pjet_i\}_{i=1}^n,k_r\bigr)\,
\\&\hphantom{=\int}\hspace*{5mm}\times
\JetRecombT\bigl(\{J_i\}_{i=1}^n\leftarrow
    \{k_i(\{\pjet_i\}_{i=1}^n,k_r)\}_{i=1}^{n+1}\bigr)
    \,
\\ &\hphantom{=\int}\hspace*{5mm}\times
\matelt{0}{n+1}\bigl(\JSum\rightarrow
   \{k_j(\{\pjet_i\}_{i=1}^n,k_r)\}_{j=1}^{n+1}\bigr)
\,.
\end{aligned}
\label{NLORealFullyDifferentialCrossSection3}
\end{equation}
The first term is finite, because any two IRCS jet algorithms will
necessarily give identical results in the
singular limit.  So long as both approach the singular limit
smoothly, this will give a convergence factor compensating
the singularities, because all divergences in Yang--Mills theory
are logarithmic.  It is important to note that this finiteness
holds point-by-point in the jet phase space so long as both
algorithms yield massless jets.  In contrast to subtraction
algorithms~\cite{CataniSeymour,FKS}, the subtraction term here is
in the jet algorithm rather than the matrix element.  It shares
the goal of isolating divergences in a simpler analytic 
structure.

We set aside the first term, to be computed numerically
after further remappings whose exploration we postpone to future
work.

The divergences of the original 
expression~(\ref{NLORealFullyDifferentialCrossSection}) are now
captured exactly by the second ($\modreal$) term.  Here, we choose the
jet algorithm to \textit{be\/} the remapping,
\begin{equation}
\JetRecombT\bigl(\{J_i\}_{i=1}^n\leftarrow
    \{k_i(\{\pjet_i\}_{i=1}^n,k_r)\}_{i=1}^{n+1}\bigr) =
    \sum_A \Theta_A(\{k_j\}_{j=1}^{n+1})\prod_{i=1}^n 
      \delta^{(D)}\bigl(J_i-\pjet_i^A(\{k_j\}_{j=1}^{n+1})\bigr)\,,
\end{equation}
with different remappings in different sectors of phase space (indexed by $A$).
We can now perform the $\pjet_i$ integrations, to obtain,
\begin{equation}
\begin{aligned}
\frac{d^{nD}\sigma^{\realT}}{\djets{D}{n}}\,&=
\int d\LIPSS^D(k_r)\,
\Jac\bigl(\{J_i\}_{i=1}^n,k_r\bigr)\,
\matelt{0}{n+1}\bigl(\JSum\rightarrow
    \{k_j(\{J_i\}_{i=1}^n,k_r)\}_{j=1}^{n+1}\bigr)
\,.
\end{aligned}
\label{NLORealFullyDifferentialCrossSectionFinal}
\end{equation}

Our choice of theory-friendly algorithm will also
require a partitioning of phase space into sectors.  That 
partitioning does not affect the decompositions we pursue
in later sections.  For simplicity's sake,
we suppress details of it in this article.

The integration in \eqn{NLORealFullyDifferentialCrossSectionFinal}
will give rise to singularities in $\eps$, but the 
KLN theorem tells us that they will cancel against the virtual 
contributions, so that the sum,
\begin{equation}
\frac{d^{nD}\sigma^{\nloT}}{\djets{D}{n}} =
\frac{d^{nD}\sigma^{\virtual}}{\djets{D}{n}}
+\frac{d^{nD}\sigma^{\realT}}{\djets{D}{n}}\,,
\label{CancelDivergent}
\end{equation}
is finite for any value of the jet momenta $J_i$.  
(The $J_i$ arguments here can of course be treated
as just dummy arguments.)
The full NLO corrections require the addition of the finite
term set aside above,
\begin{equation}
\frac{d^{nD}\sigma^{\nlo}}{\djets{D}{n}} =
\frac{d^{nD}\sigma^{\nloT}}{\djets{D}{n}}
+\frac{d^{nD}\sigma^{\differ}}{\djets{D}{n}}\,.
\label{NLOFullyDifferentialCrossSection}
\end{equation}

This quantity could then be integrated over the jet phase 
space, multiplied by
an appropriate binning delta function, in order to obtain
a prediction for the 
NLO corrections to any jet observable of interest,
\begin{equation}
\frac{d\sigma^{\rlap{$\scriptstyle\rm NLO$}{}}}{d\obs} \,= 
\int d\LIPS^D_n\bigl(K;\{J_i\}_{i=1}^n\bigr)\,
\,\delta\bigl(\obs-\Obs(\{J_i\}_{i=1}^n)\bigr)\,
\CutsTheta_n\bigl(\{J_i\}_{i=1}^n\bigr)
\frac{d^{nD}\sigma^{\rm NLO}}{\djets{D}{n}}\,.
\label{NLODifferentialCrossSection}
\end{equation}
As in~\eqn{DifferentialCrossSection}, the integration
over the jet phase space could be performed in four dimensions, and 
could be performed
numerically.

This is \textit{not\/} the way that 
subtraction~\cite{CataniSeymour,FKS} or
slicing~\cite{Slicing} methods work; the virtual-correction 
and real-emission integrals are taken
over different phase spaces, so that the large numerical 
cancellations implicit
in the sum of~\eqns{NLOVirtualDifferentialCrossSection}%
{NLORealFullyDifferentialCrossSection}
occur in the computed distributions and are a important source
of numerical uncertainty and drain on computational resources.

The matrix-element method at NLO also defines a fully differential jet 
cross section from real emission.  Approaches described in the 
literature~\cite{MEM@NLO}, however, make use of
either phase-space slicing or subtraction in order to isolate 
singular contributions, 
compute them, and cancel them against singularities in the 
virtual contributions.  

Our goal is to lay out a different path.  The integral 
in \eqn{NLORealFullyDifferentialCrossSectionFinal} is
of course divergent, so it cannot be performed numerically.  
At the same time,
it may appear analytically intractable.  Our aim in this paper is to 
take the first steps
towards rendering it tractable.

\def\Arclus{\textsc{Arclus}}
We discuss in detail the first two keys essential to pursuing 
this path.  
The first is the choice of
an analytically tractable jet algorithm.
We will make use of a $3\rightarrow 2$ clustering algorithm,
based on antenna factorization~\cite{Antenna} and harking back to 
the \Arclus{} 
algorithm~\cite{Arclus}.  
We will review the $3\rightarrow2$ clustering in the next section.  
In a real-world
computation, this will leave us with an additional 
integral as in~\eqn{NLORealFullyDifferentialCrossSection3} 
quantifying the difference
between this jet algorithm, and an experimental one (typically 
the anti-$k_{\textrm{T}}$
jet algorithm).  As mentioned
above, we leave a detailed discussion of the 
difference contribution to
future work, limiting ourselves here to the comment that it is 
finite bin-by-bin
in any distribution because of infrared safety.

The other key to a practical approach is an integrand or 
integral reduction to
a basis.  Such bases are almost universally employed in modern 
loop amplitude
calculations.   The delineation of such a basis is one of the key 
results of
this article.  We will show that a finite basis exists, and consists 
of integrands arising from products of tree-amplitude terms
with at most four distinct denominator factors.  
This is the first step to obtaining the analog of showing that
an arbitrary one-loop integrand can be written in terms 
of integrands with five or fewer propagators.  At one loop, 
one can in fact do better for integrals.
We leave a refinement of our current results to future work.

The best approach requires making a special (if by now standard)
choice of dimensional regulator.  We will discuss this choice in
\sect{DimensionalRegularizationSection}.

\section{Antenna Factorization Revisited}
\label{AntennaFactorizationSection}

\def\ah{{\hat a\vphantom{\hat b}}}
\def\bh{{\hat b}}
\def\au{{a\vphantom{\hat b}}}
\def\bu{{b\vphantom{\hat b}}}
\def\su{{s\vphantom{\hat b}}}

The $3\rightarrow2$ clustering algorithm we will use is based on 
the antenna mapping~\cite{Antenna}.  While we cannot cluster two 
massless partons into a single massless
object while maintaining overall momentum conservation in an event, we 
can cluster three
massless partons into two massless objects, which we can think 
of as protojets.
In the usual application, the trio of partons is color ordered, and 
the first and last
are `hard', while the middle parton is `singular' in the sense 
that it is soft or collinear to one of the `hard' partons.
This aspect will require restriction to a slice in phase space,
but we can cover all of phase space with appropriate slices.
We can and will use the mapping without 
reference to a color ordering.

\def\rhof{\rho}
\def\interpolate{\lambda}
\def\kf{\hat k}
If we impose parity and fix an overall $(D-1)$-dimensional rotation of the two jets
by requiring them to follow the hard-parton momenta in all singular limits, the $3\rightarrow2$
mapping is unique up to a single function.  We can map three massless parton momenta 
$i$, $j$, and $k$ to two massless protojet momenta $\ah$ and $\bh$,
\begin{equation}
\begin{aligned}
k_{\ah} &= \kf_{\ah}(k_i,k_j,k_k) = 
\\&\hphantom{=\ } \frac12\Bigl(
    1 + \rhof(i, j, k) +\frac{\rhof(i, j, k) - \interpolate(i, j, k)}
                            {s_{ik}+s_{ij}} s_{jk} \Bigr)\,k_i
    + \frac12\Bigl(1+\interpolate(i, j, k)\Bigr)\,k_j
    \\&\hphantom{=\ }
    +\frac12\Bigl(1 - \rhof(i, j, k) - 
       \frac{\rhof(i, j, k) + \interpolate(i, j, k)}{s_{ik}+s_{jk}} s_{ij}\Bigr)\,k_k\,,
\\ k_{\bh}  &= \kf_{\bh}(k_i,k_j,k_k) = 
\\&\hphantom{=\ } \frac12\Bigl(
    1 - \rhof(i, j, k) -\frac{\rhof(i, j, k) - \interpolate(i, j, k)}
                             {s_{ik}+s_{ij}} s_{jk} \Bigr)\,k_i
    + \frac12\Bigl(1-\interpolate(i, j, k)\Bigr)\,k_j
    \\&\hphantom{=\ }
    +\frac12\Bigl(1 + \rhof(i, j, k) +
       \frac{\rhof(i, j, k) + \interpolate(i, j, k)}{s_{ik}+s_{jk}} s_{ij}\Bigr)\,k_k\,.
\end{aligned}
\label{AntennaMapping}
\end{equation}
along with a singular (radiated) momentum $k_r$,
\begin{equation}
k_r = k_j\,.
\label{AntennaMapping2}
\end{equation}
We will refer to the momenta $k_i$, $k_j$ and $k_k$ as the ``recombining momenta''.

The dimensionless 
function $\interpolate(i,j,k)$ can be chosen freely, subject only to
sensible restrictions on its analytic structure and keeping the
other dimensionless function appearing, $\rhof(i,j,k)$, real.
It must be non-singular
in soft and collinear limits, though it need not have a well-defined value in the
soft limit.
With appropriate special choices, we can recover the Catani--Seymour mapping~\cite{CataniSeymour}.
The $\rhof(i,j,k)$ function is given by,
\begin{equation}
\rho(i,j,k) = \biggl[1+\frac{(1 - \interpolate^2(i,j,k)) s_{ij} s_{jk}}{K^2 s_{ik}}\biggr]^{1/2}
\,.
\label{RhoForm}
\end{equation}
In this equation, $K^2 = s_{\ah \bh} = s_{ijk}$ is the total invariant mass of the three partons
or equivalently the two protojets into which they are clustered.  We obtain \eqn{AntennaMapping}
by imposing on-shell conditions and momentum conservation to linear combinations of 
$k_{i}$, $k_j$, and $k_{k}$.
Keeping $\rhof(i,j,k)$ real implies that,
\begin{equation}
|\interpolate(i,j,k)| \le \biggl[1+\frac{K^2 s_{ik}}{s_{ij} s_{jk}}\biggr]^{1/2}\,.
\end{equation}
If we want a momentum-independent limit on $\interpolate(i,j,k)$, we must insist that,
\begin{equation}
\interpolate(i,j,k) \le 1\,.
\end{equation}

We can see that \eqn{AntennaMapping} has the correct collinear limit independent of the 
details of $\interpolate(i,j,k)$, by examining the soft limit and collinear
limits.  In the soft limit, where $k_j\rightarrow 0$,
and consequently $s_{ij}, s_{jk}\rightarrow 0$, we see that,
\begin{equation}
\rhof(i,j,k) \rightarrow 1\,,
\end{equation}
and hence,
\begin{equation}
\begin{aligned}
k_{\ah} &\rightarrow k_i\,,
\\ k_{\bh}  &\rightarrow k_k\,.
\end{aligned}
\label{AntennaMappingSoft}
\end{equation}
In the collinear limit $k_i\parallel k_j$, $s_{ij} \rightarrow 0$, and,
\begin{equation}
\begin{aligned}
k_i &\rightarrow (1-z) K_{ij}\,,
\\ k_j &\rightarrow z K_{ij}\,,
\end{aligned}
\end{equation}
where $K_{ij} = k_i+k_j$.
Accordingly, again $\rhof(i,j,k) \rightarrow 1$, and hence,
\begin{equation}
\begin{aligned}
k_{\ah} &\rightarrow \bigl(
    1-z +\frac{z}2(1 - \interpolate(i, j, k))\bigr)\,K_{ij}
    + \frac{z}2\bigl(1+\interpolate(i, j, k)\bigr)\,K_{ij}
    \\ &= K_{ij} \,,
\\ k_{\bh}  &\rightarrow -\frac{z}2\bigl(1-\interpolate(i, j, k)\bigr)\,K_{ij}
    + \frac{z}2\bigl(1-\interpolate(i, j, k)\bigr)\,K_{ij}
    +k_k
    \\ &= k_k\,.
\end{aligned}
\label{AntennaMappingCollinear}
\end{equation}
A similar simplification arises in the other collinear limit, where $k_j\parallel k_k$.

The explicit solutions in \eqn{AntennaMapping} contain square roots 
and rational
functions. These are not compatible with the 
language of computational algebraic
geometry we will employ in later sections.  That language wants 
a reformulation
in polynomial form.
As an example of the class of reformulations we will use, let us
re-express the constraints in purely polynomial form.  Writing,
\begin{equation}
\begin{aligned}
k_{\ah} &= c_{\ah,i} k_i + c_{\ah,j} k_j + c_{\ah,k} k_k\,,
\\
k_{\bh} &= c_{\bh,i} k_i + c_{\bh,j} k_j + c_{\bh,k} k_k\,,
\end{aligned}
\label{JetMappingGeneral}
\end{equation}
we have,
\begin{equation}
\begin{aligned}
k_{\ah}^2 &= 0 = c_{\ah,i} c_{\ah,k} s_{ik} + c_{\ah,i} c_{\ah,j} s_{ij} + 
   c_{\ah,k} c_{\ah,j} s_{jk}\,, 
\\
k_{\bh}^2 &= 0 = c_{\bh,i} c_{\bh,k} s_{ik} + c_{\bh,i} c_{\bh,j} s_{ij} + 
   c_{\bh,k} c_{\bh,j} s_{jk}\,,
\end{aligned}
\label{eq:ProtojetOnShell}
\end{equation}
for the protojet on-shell conditions, and
\begin{equation}
\begin{aligned}
0 &= (-1 + c_{\ah, k} + c_{\bh, k}) s_{ik} + (-1 + c_{\ah, j} + c_{\bh, j}) s_{ij}\,, 
\\
0 &= (-1 + c_{\ah, i} + c_{\bh, i}) s_{ik} + (-1 + c_{\ah, j} + c_{\bh, j}) s_{jk}\,, 
\\
0 &= (-1 + c_{\ah, i} + c_{\bh, i}) s_{ij} + (-1 + c_{\ah, k} + c_{\bh, k}) s_{jk}\,,
\end{aligned}
\label{eq:ProtojetMomCons}
\end{equation}
as momentum conservation conditions.  We can simplify these equations by solving linear ones,
\begin{equation}
c_{\bh,x} = 1-c_{\ah,x}\,.
\end{equation}
Combined, \eqns{eq:ProtojetOnShell}{eq:ProtojetMomCons} are 5 equations in 6 unknowns (the $c_{\ah}$ and $c_{\bh}$). It is easy to see that the remaining degree of freedom can be identified with $\lambda$. Specifically, 
we can identify $c_{\ah,j} = (1+\lambda(i,j,k))/2$. This leaves 
us with two coefficients
($c_{\ah,i}$ and $c_{\ah,k}$) to be
treated later as additional independent variables, alongside
a pair of equations,
\begin{equation}
\begin{aligned}
0 &= (c_{\ah, i} + c_{\ah, k}) s_{ik} 
+ (c_{\ah, i} + {\textstyle\frac12}(1+\lambda(i,j,k))) s_{ij} 
    + (c_{\ah, k} + {\textstyle\frac12}(1+\lambda(i,j,k))) s_{jk}
    -s_{\ah\bh} \,,
\\
0 &= 2 c_{\ah, i} c_{\ah, k} s_{ik} 
+ c_{\ah, i} (1+\lambda(i,j,k)) s_{ij} 
     + c_{\ah, k} (1+\lambda(i,j,k)) s_{jk}\,.
\end{aligned}
\label{CoefficientEquationPair}
\end{equation}
We note that the linear equation can also be solved, but only in an asymmetric manner; we avoid doing this.
Substituting the solutions into \eqn{JetMappingGeneral}, we obtain
a compact polynomial representation for the mapping,
\begin{equation}
\begin{aligned}
k_{\ah} &= c_{\ah,i} k_i + {\textstyle\frac12}(1+\lambda(i,j,k)) k_j + c_{\ah,k} k_k\,,
\\
k_{\bh} &= (1-c_{\ah,i}) k_i + {\textstyle\frac12}(1-\lambda(i,j,k)) k_j + (1-c_{\ah,k}) k_k\,.
\end{aligned}
\label{JetMappingCompact}
\end{equation}

We will also take the $\lambda$ function as an additional 
independent variable.

Let us define two ``remapping constraint functions'' corresponding to \eqn{CoefficientEquationPair},
\begin{equation}
\begin{aligned}
R_1(c_1,c_2,s_1,s_2,s_3,\lambda) &\equiv 
2 (c_1+c_2) s_1 + (2c_1 + (1+\lambda)) s_2 
   \\&\hphantom{\equiv!} + (2c_2 + (1+\lambda)) s_3-2(s_1+s_2+s_3) \,,
\\
R_2(c_1,c_2,s_1,s_2,s_3,\lambda) &\equiv 
2 c_1 c_2 s_1 + c_1 (1+\lambda) s_2 + c_2 (1+\lambda) s_3\,.
\end{aligned}
\label{RemappingConstraint}
\end{equation}
Because the solutions for $c_{\ah,i}$ and $c_{\ah,k}$ are algebraic
rather than polynomial
in terms of the invariants, we do not want to solve for them 
explicitly.
We could use them to impose the constraints of \eqn{CoefficientEquationPair} in the form
of equations,
\begin{equation}
    \begin{aligned}
    R_1&(c_{\ah,i},c_{\ah,k},s_{ik},s_{ij},s_{jk},\lambda) = 0\,,
    \\R_2&(c_{\ah,i},c_{\ah,k},s_{ik},s_{ij},s_{jk},\lambda) = 0\,.
    \end{aligned}
\end{equation}

With given four-momentum $k_{\ah}+k_{\bh}=k_i+k_j+k_k$ dumped into 
the triplet of partons $\{i,j,k\}$,
the phase-space boundaries are determined by the nonnegativity of parton energies along with
the usual Dalitz-triangle constraint~\cite{BycklingKajantie} $s_{ij}+s_{jk}\le K^2$.
(We need a third variable to fully parametrize the partonic phase space, but it does not
control the singular behavior of matrix elements.)  In the remapped variables, the boundary
constraints take a very similar form,
\begin{equation}
s_{\ah r} \ge 0\,,\quad 
s_{r \bh} \ge 0\,,\quad
s_{\ah r} + s_{r \bh} \le K^2\,.
\label{RemappedPhaseSpaceBoundaries}
\end{equation}

\section{Dimensional Regularization}
\label{DimensionalRegularizationSection}

The expressions in \sect{JetObservablesSection} 
make use of dimensional regularization.  All integrations
are in $D=4-2\eps$ dimensions.  This is of course not 
suitable for the ultimate numerical
integrations we must perform, so we must take the four-dimensional 
limit of appropriate
intermediate expressions.  In the LO truncation of 
\eqn{DifferentialCrossSection},
the fully differential cross section is finite as $\eps\rightarrow 0$, 
so in that expression we can take the limit.
The jet cuts eliminate regions where it has singularities, 
so we can also restrict
the integration to four dimensions, and take $\eps\rightarrow 0$ in 
the measure as well.
The extra integrations would contribute corrections of $\Ord(\eps)$ to 
physical observables;
but at the end of a calculation, we can certainly take 
$\eps\rightarrow 0$, whereupon they will vanish.
We can then calculate the tree-level matrix element in 
\eqn{LOFullyDifferentialCrossSectionFinal} 
using four-dimensional techniques, which are critical to efficiency.

At NLO, the situation is more subtle.  The virtual matrix 
element contains explicit poles
in $\eps$, so we cannot take the limit there. While 
the $\eps$-dimensional integrations
over the corresponding components of the
jet momenta are not needed in order to regulate singularities, they 
would shift the results
for differential cross sections
by finite amounts.
In a physical quantity, this would be canceled by an opposite shift in 
the real-emission
contributions.  The treatment in \sect{JetObservablesSection} 
corresponds to the traditional,
``conventional'' (CDR) scheme in which all momenta and spins 
are continued uniformly to
$4-2\eps$ dimensions.  A different approach keeps observable momenta 
in four dimensions,
and continues only unphysical momenta --- virtual particles, soft 
or collinear real emissions ---
to $4-2\eps$ dimensions.  This approach, common to the 
four-dimensional helicity~(FDH), 
't~Hooft--Veltman~(HV), and dimensional reduction~(DR) schemes, 
makes integral bases smaller.
It also allows the use of four-dimensional techniques 
in computing loop
amplitudes.  We shall denote these variants of dimensional 
regularization ``four-dimensional''
schemes below.  They replace 
\eqn{NLOVirtualDifferentialCrossSection} by,
\begin{equation}
\frac{d\sigma^{\rlap{$\scriptstyle\virtual$}{}}}{d\obs} = 
\int d\LIPS^4_n\bigl(K;\{J_i\}_{i=1}^n\bigr)\,
\,\delta\bigl(\obs-\Obs(\{J_i\}_{i=1}^n)\bigr)\,
\CutsTheta_n\bigl(\{J_i\}_{i=1}^n\bigr)
\frac{d^{4n}\sigma^{\virtual}}{\djets{4}{n}}\,.
\label{NLOVirtualDifferentialCrossSection4D}
\end{equation}

What is the correct real-emission analog of this virtual 
differential cross section?  It
is clear that integrals over parton momenta other than $k_i$, $k_j$, 
and $k_k$ should be done in four
dimensions; and clear that the integral over $k_r=k_j$ should be done 
in $4-2\eps$ dimensions.
 We are left to determine the correct treatment of 
the protojets $k_{\ah}$ and $k_{\bh}$ and of the partons $k_i$ 
and $k_k$: are
there additional integrals that must be included along with 
$d\LIPSS^D$?  Because of momentum
conservation, it is clear that at least one of $k_i$ and $k_k$ must 
have $\eps$-dimensional
components; but are we obliged to integrate over them?

To gain insight into the correct prescription for matching the 
four-dimensional treatment of
external momenta in the virtual contributions, we can examine the 
treatment in the slicing~\cite{Slicing} and Catani--Seymour 
subtraction~\cite{CataniSeymour} approaches.
We can also examine the potentially singular denominators that 
may arise.  In both cases,
there is effectively only \textit{one\/} degree of freedom 
regulating the singularities.
Both calculations, for example, can be embedded in five dimensions 
and do not require
a six-dimensional embedding space.

Furthermore, if we examine the singular terms in the fully
differential cross section ${d^{4n}\sigma^{\virtual}}/{\djets{4}{n}}$
in the integrand of \eqn{NLOVirtualDifferentialCrossSection4D}, we
will find that they are proportional to the four-dimensional 
leading-order fully differential cross section rather than the 
$D$-dimensional one.  Unitarity requires
that we match this in the real-emission cross section after 
integration over the singular emission.  

The different arguments above all consistently imply that the 
protojets $\ah$ and $\bh$ must
be treated as four-dimensional, and must be integrated over four 
dimensions.  Moreover,
it is most natural to take the $\eps$-dimensional components 
of $k_r=k_j$ as the 
regulating integration variables.  Momentum conservation means that 
the $\eps$-dimensional
components of $k_i$ and $k_k$ cannot vanish; but this 
degree-of-freedom restriction implies
that they cannot be independent of $k_j$ either.  Each component 
must be linearly dependent
on the corresponding component of $k_j$, so that the trio of partons 
can be embedded in five dimensions.  

\def\fourd#1#2{{\bar #1}_{#2}}
\def\epsd#1#2{\mu_{#2}}
The mapping in \eqn{AntennaMapping} is valid when either all momenta 
are strictly four-dimensional,
or when all are $D$-dimensional.  Let us now see how to modify it for 
four-dimensional regulator
schemes.  Introduce a notation for four- and $\eps$-dimensional 
components respectively,
\begin{equation}
k_i = \fourd{k}{i} + \epsd{k}{i}\,.
\end{equation}

By design, $\epsd{k}{\ah} = 0 = \epsd{k}{\bh}$; we will implement this
via additional delta functions.  This will give two linear 
equations relating
$\epsd{k}{i}$, $\epsd{k}{k}$, and $\epsd{k}{j}$.  That will force both
$\epsd{k}{i}$ and $\epsd{k}{k}$ to be proportional to $\epsd{k}{j}$, 
and allows us to solve for the former,
\begin{equation}
\begin{aligned}
\epsd{k}{i} &= 
  \frac{(c_{\ah,k}-c_{\ah,j})}{c_{\ah,i}-c_{\ah,k}} \epsd{k}{j}\,,
\\
\epsd{k}{k} &= 
  -\frac{(c_{\ah,i}-c_{\ah,j})}{c_{\ah,i}-c_{\ah,k}} \epsd{k}{j}\,.
\end{aligned}
\label{EpsMomentumSolutions}
\end{equation}
Putting in the solutions for $c_{\ah,\{i,j,k\}}$, we find,
\begin{equation}
\begin{aligned}
\epsd{k}{i} &= 
-\frac{(s_{ik} + s_{ij}) (\rho(i, j, k) + \lambda(i, j, k))}
 {(\rho(i, j, k) (K^2+s_{ik}) + (s_{ij} - s_{jk}) \lambda(i, j, k))}
                     \epsd{k}{j} \,,
\\
\epsd{k}{k} &= 
-\frac{(s_{ik} + s_{jk}) (\rho(i, j, k) - \lambda(i, j, k))}
 {(\rho(i, j, k) (K^2+s_{ik}) + (s_{ij} - s_{jk}) \lambda(i, j, k))}
                     \epsd{k}{j} \,,
\end{aligned}
\end{equation}
\Eqn{AntennaMapping} is unchanged, though it effectively takes 
the form,
\begin{equation}
\begin{aligned}
k_{\ah} &= \frac12\Bigl(
    1 + \rhof(i, j, k) +\frac{\rhof(i, j, k) - \interpolate(i, j, k)}
                            {s_{ik}+s_{ij}} s_{jk} \Bigr)\,\fourd{k}{i}
    + \frac12\Bigl(1+\interpolate(i, j, k)\Bigr)\,\fourd{k}{j}
    \\&\hphantom{=\ }
    +\frac12\Bigl(1 - \rhof(i, j, k) - 
       \frac{\rhof(i, j, k) + 
       \interpolate(i, j, k)}{s_{ik}+s_{jk}} 
       s_{ij}\Bigr)\,\fourd{k}{k}\,,
\\ k_{\bh} &= 
\frac12\Bigl(
    1 - \rhof(i, j, k) -\frac{\rhof(i, j, k) - \interpolate(i, j, k)}
                             {s_{ik}+s_{ij}} s_{jk} 
                             \Bigr)\,\fourd{k}{i}
    + \frac12\Bigl(1-\interpolate(i, j, k)\Bigr)\,\fourd{k}{j}
    \\&\hphantom{=\ }
    +\frac12\Bigl(1 + \rhof(i, j, k) +
       \frac{\rhof(i, j, k) + \interpolate(i, j, k)}{s_{ik}+s_{jk}} 
       s_{ij}\Bigr)\,\fourd{k}{k}\,.
\end{aligned}
\label{AntennaMapping4D}
\end{equation}

We can recast \eqn{EpsMomentumSolutions} as a pair
of polynomial equations in scalar variables by contracting
both sides of each equation with $\epsd{k}{j}$, and clearing 
denominators.  
 This gives us
two additional remapping constraint functions,
\begin{equation}
\begin{aligned}
R_3(c_1,c_2,e_1,e_2,e_3,\lambda) &\equiv 
(c_1-c_2) e_2 - (c_2-(1+\lambda)/2) e_1\,,
\\
R_4(c_1,c_2,e_1,e_2,e_3,\lambda) &\equiv e_1+e_2+e_3\,.
\end{aligned}
\label{RemappingConstraint4D}
\end{equation}
(The latter is obtained from the sum of the two equations in 
\eqn{EpsMomentumSolutions}.)
These two functions will also give constraints in the form,
\begin{equation}
    \begin{aligned}
     R_3(c_{\ah,i},c_{\ah,k},\mu_j^2,\mu_i\cdot\mu_j,\mu_k\cdot\mu_j,
         \lambda) &= 0\,,\\
     R_4(c_{\ah,i},c_{\ah,k},\mu_j^2,\mu_i\cdot\mu_j,\mu_k\cdot\mu_j,
         \lambda) &= 0\,.
    \end{aligned}
\end{equation}

\section{Inverse Antenna Mapping}
\label{InverseAntennaMappingSection}

\def\ft{{\tilde f}}
\def\lamt{{\tilde\lambda}}
\def\rhot{{\tilde\rho}}
\def\kinv{\tilde k}
\def\sahr{s_{\ah r}}
\def\sbhr{s_{r\bh}}
Unlike most reconstruction steps in jet algorithms, 
the antenna mapping is invertible.
Indeed, we need the inverse mapping, for example,
in order to change variables from the partonic
momenta to the protojet momenta, or equivalently from 
partonic invariants to protojet invariants.  We
can solve for $k_i$, $k_j$, and $k_k$ in terms of $k_\ah$, $k_r$, 
and $k_\bh$, as follows,
\begin{equation}
\hspace*{-6mm}\begin{aligned}
k_i &= \frac12 \bigl(1 + \tau(\sahr,\sbhr)
     w_+(\sahr,\sbhr)\bigr)\,k_\ah
-\frac12\bigl(1+\tau(\sahr,\sbhr)w_\lambda(\sahr,\sbhr)\bigr)\, k_r
    \\&\hphantom{=\ }
+\frac12 \bigl(1 + \tau(\sahr,\sbhr)
    w_-(\sahr,\sbhr)\bigr)\,k_\bh\,,
\\ k_j &= k_r\,,
\\
k_k &=
 \frac12 \bigl(1 - \tau(\sahr,\sbhr)
    w_+(\sahr,\sbhr)\bigr)\,k_\ah
-\frac12\bigl(1-\tau(\sahr,\sbhr)w_\lambda(\sahr,\sbhr)\bigr)\, k_r
   \\&\hphantom{=\ }
+\frac12 \bigl(1 - \tau(\sahr,\sbhr)
    w_-(\sahr,\sbhr)\bigr)\,k_\bh\,,
\end{aligned}
\label{InverseAntennaMapping1}
\end{equation}
where we define,
\begin{equation}
\begin{aligned}
w_0(s_1,s_2) &\equiv 1 - \frac{s_1}{K^2} 
  - \frac{s_2}{K^2} \,,
\\      
w_\lambda(s_1,s_2) &\equiv \lamh(s_1,s_2)
 + \frac{s_1}{2K^2}(1 - \lamh(s_1,s_2))  
 -\frac{s_2}{2K^2}(1+\lamh(s_1,s_2)) \,,
\\
w_\Sigma(s_1,s_2) &\equiv
      \frac{s_1}{2K^2}(1+\lamh(s_1,s_2))
      - (1-\lamh(s_1,s_2))\frac{s_2}{2K^2} \,,
\\
w_\pm(s_1,s_2) &\equiv w_\Sigma(s_1,s_2)\pm w_0(s_1,s_2)\,,
\\
w_J(s_1,s_2) &\equiv 1 - \frac{s_1+s_2}{2K^2} + \frac{s_1-s_2}{2K^2}\lamh(s_1,s_2)\,,
\end{aligned}
\label{Wdefinitions}
\end{equation}
and take $\tau(s_1,s_2)$ to be,
\begin{equation}
\begin{aligned}
\tau(s_1,s_2) = 2 K^2\,\bigl[&2(1+\lamh(s_1,s_2)) (K^2 - s_2)^2 
            + 2(1 - \lamh(s_1,s_2)) (K^2 - s_1)^2
\\ & -(1 - \lamh{}^2(s_1,s_2)) (s_1 + s_2)^2\big]^{-1/2}\,.
\end{aligned}
\label{TauValue}
\end{equation}

It is obvious in \eqn{InverseAntennaMapping1} that
$k_i+k_j+k_k = k_\ah+k_\bh$ and that $k_j^2 = 0$; the reader
may verify with a little bit of algebra that $k_i^2 = 0 = k_k^2$
as well.  These statements hold independently of the precise
form of $\lamh$.
The precise relation between the forward
mapping's $\lambda(i,j,k)$ and 
$\lamh(s_1,s_2)$ is complicated; but
if we care about the inverse mapping generally (for example, inside 
an integral) rather than
the exact inverse to \eqn{AntennaMapping}, we can choose 
the dimensionless function $\lamh$
freely, subject to sensible constraints on its analytic behavior.  

Let us verify the behavior
in the soft and collinear limits.  In the soft 
limit, $k_r,\sahr,\sbhr\rightarrow 0$; consequently,
\begin{equation}
\begin{aligned}
w_\pm(\sahr,\sbhr) &\rightarrow \pm 1\,,
\\
\tau(\sahr,\sbhr) &\rightarrow 1\,,
\end{aligned}
\end{equation}
and the inverse mapping takes the form,
\begin{equation}
\begin{aligned}
k_i &\rightarrow k_\ah\,,
\\ k_j &\rightarrow 0\,,
\\ k_k &\rightarrow k_\bh\,.
\end{aligned}
\label{InverseAntennaMappingSoft}
\end{equation}

\def\Kahr{K_{\ah\backslash r}}
In the $k_{\ah}\parallel k_r$ limit, we may 
define $\Kahr=k_{\ah}-k_r$ and then write,
\begin{equation}
\begin{aligned}
k_{\ah} &= (1+\zeta) \Kahr\,,
\\ k_r &= \zeta \Kahr\,,
\end{aligned}
\end{equation}
Here, $\sahr\rightarrow 0$ while 
$\sbhr\rightarrow \zeta K^2/(1+\zeta)$, and accordingly,
\begin{equation}
\begin{aligned}
w_0(\sahr,\sbhr) &\rightarrow \frac{1}{1+\zeta} \,,
\\      
w_\lambda(\sahr,\sbhr) &\rightarrow \lamh(\sahr,\sbhr)
 -\frac{\zeta}{2(1+\zeta)}(1+\lamh(\sahr,\sbhr)) \,,
\\
w_\Sigma(\sahr,\sbhr) &\rightarrow
      - \frac{\zeta}{2(1+\zeta)}(1-\lamh(\sahr,\sbhr)) \,,
\\
w_\pm(\sahr,\sbhr) &\rightarrow 
\frac{\pm2-\zeta}{2(1+\zeta)}
      +\lamh(\sahr,\sbhr)\frac{\zeta}{2(1+\zeta)} \,,
\\
\tau(\sahr,\sbhr) &\rightarrow \frac{2(1+\zeta)}{2+(1-\lamh(\sahr,\sbhr))\zeta}\,.
\end{aligned}
\label{Wcollinear}
\end{equation}
We then find for the inverse mapping,
\begin{equation}
\begin{aligned}
k_i &\rightarrow \Kahr\,,
\\ k_j &\rightarrow \zeta\Kahr\,,
\\ k_k &\rightarrow k_{\bh}\,.
\end{aligned}
\label{InverseAntennaMappingCollinear}
\end{equation}

For the Jacobian for the change of variables from $(k_i,k_j,k_k)$ to
$(k_\ah,k_r,k_\bh)$, we obtain (see appendix~\ref{JacobianAppendix}
for details),
\begin{equation}
\begin{aligned}
\frac{d^D k_i}{(2\pi)^4}&\deltap(k_i^2)\,
\frac{d^D k_j}{(2\pi)^4}\deltap(k_j^2)\,
\frac{d^D k_k}{(2\pi)^4}\deltap(k_k^2) =
\\&
\tau^{D-1}(s_{\ah r},s_{r\bh}) w_0^{D-3}(s_{\ah r},s_{r\bh})
\\&\times
\biggl[
 w_J(s_{\ah r},s_{r\bh})
+ \frac{s_{\ah r}s_{r\bh}}{K^2}\Bigl[
 \frac{\partial\lamh(s_{\ah r},s_{r\bh})}{\partial s_{\ah r}}
   -\frac{\partial\lamh(s_{\ah r},s_{r\bh})}{\partial s_{r\bh}}
   \Bigr]\biggr]
\\&\times
\frac{d^D k_\ah}{(2\pi)^4}\deltap(k_\ah^2)\,
\frac{d^D k_\bh}{(2\pi)^4}\deltap(k_\bh^2)\,
\frac{d^D k_r}{(2\pi)^4}\deltap(k_r^2)\,.
\end{aligned}
\label{InverseAntennaJacobian}
\end{equation}

The only singularities appearing here are associated with $\tau$, and
this is also true when changing variables in invariants,
\begin{equation}
\begin{aligned}
s_{ij} = {}
 &\frac{s_{\ah r}}2 \bigl(1 + \tau(\sahr,\sbhr) w_+(\sahr,\sbhr)\bigr)
+\frac{s_{r\bh}}2 \bigl(1 + \tau(\sahr,\sbhr) w_-(\sahr,\sbhr)\bigr)
\,,
\\
s_{jk} = {}
 &\frac{s_{\ah r}}2 \bigl(1 - \tau(\sahr,\sbhr) w_+(\sahr,\sbhr)\bigr)
+\frac{s_{r\bh}}2 \bigl(1 - \tau(\sahr,\sbhr) w_-(\sahr,\sbhr)\bigr)
\,,
\\
s_{ik} = {}&K^2-s_{\ah r}-s_{r\bh}\,,
\\
s_{i x} = {}
&\frac12 s_{\ah x}
\bigl(1 + \tau(\sahr,\sbhr) w_+(\sahr,\sbhr)\bigr)
  +\frac12 s_{\bh x} 
     \bigl(1 + \tau(\sahr,\sbhr) w_-(\sahr,\sbhr)\bigr) 
\\&-\frac12 s_{r x}
\bigl( 1 +\tau(\sahr,\sbhr) w_\lambda(\sahr,\sbhr)\bigr)\,,
\\
s_{k x} = {}
&\frac12s_{\ah x}
\bigl(1 - \tau(\sahr,\sbhr) w_+(\sahr,\sbhr)\bigr)
  +\frac12 s_{\bh x} 
     \bigl(1 - \tau(\sahr,\sbhr) w_-(\sahr,\sbhr)\bigr) 
\\&-\frac12 s_{r x}
\bigl( 1 - \tau(\sahr,\sbhr) w_\lambda(\sahr,\sbhr)\bigr)\,.
\end{aligned}
\label{InvariantsInverse}
\end{equation}
In these equations, $x$ is a massless momentum other than $(i,j,k)$, and so necessarily
four-dimensional.

\def\tc{\tilde c}
As in the forward mapping, we will need polynomial expressions for the 
constraints, avoiding
the square root in \eqn{TauValue}.  Write,
\begin{equation}
\begin{aligned}
k_i &= \tc_{i,\ah} k_{\ah} + \tc_{i,r} k_r + \tc_{i,\bh} k_{\bh}\,,
\\ k_k &= \tc_{k,\ah} k_{\ah} + \tc_{k,r} k_r + \tc_{k,\bh} k_{\bh}\,,
\end{aligned}
\end{equation}
alongside $k_j=k_r$.  Momentum conservation requires that,
\begin{equation}
\begin{aligned}
\tc_{k,\ah} &= 1-\tc_{i,\ah}\,,
\\ \tc_{k,\bh} &= 1-\tc_{i,\bh}\,,
\\ \tc_{k,r} &= -1-\tc_{i,r}\,.
\end{aligned}
\label{MomentumConservationCoefficients}
\end{equation}
We also obtain the constraints,
\begin{equation}
\begin{aligned}
k_i^2 &= 0 = 
K^2 \tc_{i,\ah} \tc_{i,\bh} + s_{\ah r} \tc_{i,\ah} \tc_{i,r} 
  + s_{r\bh} \tc_{i,\bh} \tc_{i,r}
\,,
\\ k_k^2 &= 0 = 
K^2 (1 - \tc_{i,\ah}) (1 - \tc_{i,\bh})
-s_{\ah r} (1-\tc_{i,\ah}) (1+\tc_{i,r})
-s_{r\bh} (1-\tc_{i,\bh}) (1+\tc_{i,r})
\,.
\end{aligned}
\end{equation}
We can rearrange these into a linear and a quadratic constraint,
\begin{equation}
\begin{aligned}
0 &= 
K^2 (1-\tc_{i,\ah}- \tc_{i,\bh})
  -s_{\ah r} (1 - \tc_{i,\ah} + \tc_{i,r})
  -s_{r\bh} (1-\tc_{i,\bh}+\tc_{i,r})
\,,
\\ 0 &= 
K^2 \tc_{i,\ah} \tc_{i,\bh} + s_{\ah r} \tc_{i,\ah} \tc_{i,r} 
+ s_{r\bh} \tc_{i,\bh} \tc_{i,r}
\,.
\end{aligned}
\label{InverseMappingCoefficientConstraints}
\end{equation}
In terms of $\tau$ and $\lamh$, the first equation is 
satisfied trivially,
while upon removing an overall factor of 
$(K^2-s_{\ah r}-s_{r\bh})/(8 (K^2)^2)$,
the second one becomes,
\begin{equation}
\begin{aligned}
0 &= 
4 (K^2)^2 - \bigl[4 (K^2)^2 - 4 K^2 s_{\ah r}(1-\lamh(s_{\ah r},s_{r\bh}))
  - 4 K^2 s_{r\bh}(1+\lamh(s_{\ah r},s_{r\bh}))
\\ &\hspace*{20mm}
+\bigl((1+\lamh(s_{\ah r},s_{r\bh}))s_{r\bh}
  -(1-\lamh(s_{\ah r},s_{r\bh}))s_{\ah r}\bigr)^2
\bigr]\,\tau^2(s_{\ah r},s_{r\bh})\,.
\end{aligned}
\label{TauEquation}
\end{equation}
Its solution is \eqn{TauValue}.  For later use, we define the
right-hand side as a constraint function,
\def\Rtau{R_\tau}
\begin{equation}
\begin{aligned}
\Rtau(s_{\ah r},s_{r\bh},\tau,\lamh) &\equiv     
4 (K^2)^2 - \bigl[4 (K^2)^2 - 4 K^2 s_{\ah r}(1-\lamh)
  - 4 K^2 s_{r\bh}(1+\lamh)
\\ &\hspace*{20mm}
+\bigl((1+\lamh)s_{r\bh}
  -(1-\lamh)s_{\ah r}\bigr)^2
\bigr]\,\tau^2\,.
\end{aligned}
\label{TauConstraint}
\end{equation}

\section{One-Loop Decomposition Revisited}
\label{OneLoopDecompositionSection}

Our goal in the present paper is to show that the integrand
arising from any squared tree-level matrix element 
$\matelt{0}{n+1}$ can be written
in terms of a (small) set of master integrands, and to delineate
such a set.  We leave the determination of a minimal set to
future work.

We will recast the reduction of one-loop integrands
for an $n$-point process to simpler integrands in a technology
which we can apply to the problem of reducing real-emission
integrands compatible with the theoretical jet algorithm
introduced in \sect{JetObservablesSection} and specified
in \sects{AntennaFactorizationSection}{InverseAntennaMappingSection}.

In modern one-loop amplitude calculations, one makes use of
a standard integral basis, in which any Feynman integral
in a process of interest can be written as a sum of
basis or `master' integrals with rational coefficients 
(rational in the Lorentz invariants
and $\eps$).  
Such a decomposition is relevant even if we do not make use of
unitarity in computing the coefficients of the integrals in
the expression for the amplitude.

We first review the initial steps in deriving this basis,
focusing on the integrands rather than on the integrals.
The well-known derivation relies on two techniques:
partial fractioning, and use of dimensional identities.
There are two reductions that we must consider, of $n$-point
integrands to lower-point integrands down to a stopping
point; and of reductions of integrands with nontrivial numerators to
integrands with simpler numerators.

\def\proj{\flat}
The integrand in each Feynman loop integral will have numerators 
build out of
Lorentz products of the loop momenta $\{\ell_i\}$, external 
momenta $\{k_i\}$, and other external
vectors $\{w_i\}$ (such as spinor strings or polarization vectors).  
We work in a variant of dimensional regularization where all external
momenta and vectors are strictly four-dimensional.
The denominators are propagator
denominators, with power unity except for cases involving bubbles 
on internal lines.  The external vectors ($\{w_i\}$) 
in the numerator can be written in terms of a basis $\{v_i\}$; 
if there are more than four independent momentum arguments to an 
integral, external momenta suffice to form a basis.  This is the case
on which we focus first, corresponding to Feynman integrals
with five or more external legs.  We take all internal
lines to be massless, while external legs can be either
massless or massive.

Consider first consider the reduction of `scalar' $n$-point 
integrands (\textit{i.{}e.{}\/} whose
numerator is independent of the loop momentum) to simpler
integrals.  This is in a sense the most
conceptually important reduction, as it takes us from a 
potentially infinite set of basis
integrals to a finite set.

We will make use of Gram determinants,
\begin{equation}
G\bigg(\gatop{p_1,\ldots,p_m}{q_1,\ldots,q_m}\bigg) =
 \det_{i,j}\big(2p_i\cdot q_j\big)\,.
\end{equation}
When the sequences $p_1,\ldots p_m$ and $q_1,\ldots,q_m$ are
identical, we list only one,
\begin{equation}
G\bigl({p_1,\ldots,p_m}\bigr) \equiv 
G\bigg(\gatop{p_1,\ldots,p_m}{p_1,\ldots,p_m}\bigg)
\end{equation}
If we consider an integrand with $n>5$, then we have $5$ 
formally independent momenta,
and we can make use of the following Gram determinant,
\begin{equation}
G\biggl(\,
\begin{matrix}
\ell,\!&\!k_1,\!&\!\ldots,\!&\!k_{4}\\[-1mm]
k_{5},\!&\!k_1,\!&\!\ldots,\!&\!k_{4}
\end{matrix}
\biggr)\,.
\end{equation}
It evaluates to zero for all values of loop momentum, 
because of the linear dependence 
between the five momenta induced by the dimension of space-time.  
This vanishing is of course
not manifest, and thus gives rise to non-trivial relations 
between Lorentz invariants
involving the loop momentum.  Equivalently, it gives rise 
to non-trivial relations between
different denominators.  These nontrivial relations are what 
make partial-fractioning
the integrand possible.  It is precisely the analog of these 
relations that we will
explore in later sections for real-emission integrals.

\section{Reduction of Numerator-Free Loop Integrands}
\label{LoopNumeratorFreeReductionSection}

Let us begin by revisiting the partial fractioning of integrands 
of scalar one-loop 
integrals\footnote{We use `scalar' to mean `numerator free of the 
loop momentum'}. 
As an example, consider the integrand of the six-point 
one-loop integral with external
momenta in four dimensions and all
legs massless.  The denominators are,
\begin{equation}
D_j = \bigl(\ell- K_{1,j}\bigr)^2\,,
\label{DenominatorForm}
\end{equation}
where,
\begin{equation}
K_{j,l} \equiv \sum_{i=j}^{l-1} k_i\,.
\end{equation}
Empty sums are understood to vanish, and the index runs cyclicly 
mod~6.  We can then rewrite,
\begin{equation}
G_1 \equiv G\biggl(\,
\begin{matrix}
\ell,\!&\!k_1,\!&\!\ldots,\!&\!k_{4}\\[-1mm]
k_{5},\!&\!k_1,\!&\!\ldots,\!&\!k_{4}
\end{matrix}
\biggr) = \omega_j D_j + \omega_0\,,
\label{GramDecomposition}
\end{equation}
in which $j$ is implicitly summed over $1\ldots6$.

\def\Xed#1{\xcancel{#1}}
In this expression,
\begin{equation}
\begin{aligned}
\omega_1 &= -G\biggl(
\begin{matrix}
\,k_1\!&\!k_2,\,k_3,\!&\!k_4\\[-1mm]
\,k_{5},\!&\!k_2,\,k_3,\!&\!k_{4}
\end{matrix}
\biggr)\,,
\\ \omega_{2\le j\le5} &= G\biggl(
\begin{matrix}
\,k_1\!&\!k_2,\,\,k_3,\!&\!k_4\\[-1mm]
\,k_{j+1},\!&\!\ldots\Xed{k_6}\ldots,\!&\!k_{j-1}+k_j
\end{matrix}
\biggr)\,,
\\ \omega_6 &= -G\biggl(
\begin{matrix}
\,k_1\!&\!k_2,\,k_3,\!&\!k_4\\[-1mm]
\,k_{1},\!&\!k_2,\,k_3,\!&\!k_{4}
\end{matrix}
\biggr)\,,
\end{aligned}
\end{equation}
where the cross-out indicates an omitted momentum.

The inhomogeneous term in \eqn{GramDecomposition} is,
\begin{equation}
\omega_0 = -\omega_j K_{1,j}^2\,.
\end{equation}
This appears to be generally positive, as evaluated on
randomly chosen configurations of momenta; but we offer no
proof that it avoids vanishing.  We will assume that it is nonzero.

Using \eqn{GramDecomposition}, we can then write the 
following identity,
\begin{equation}
\frac{G_1}{D_1 D_2 D_3 D_4 D_5 D_6} = 0 = 
\sum_{j=1}^6\frac{\omega_j}{D_1\cdots \Xed{D_j}\cdots D_6} 
+ \frac{\omega_0}{D_1 D_2 D_3 D_4 D_5 D_6}\,,
\label{DenominatorIdentity}
\end{equation}
where the cross-out indicates an omitted denominator.
This gives a partial-fractioning identity,
\begin{equation}
\frac{1}{D_1 D_2 D_3 D_4 D_5 D_6}
=
-\sum_{j=1}^6\frac{\omega_j/\omega_0}
     {D_1\cdots \Xed{D}_j\cdots D_6} \,.
\label{DenominatorIdentityII}
\end{equation}

Upon integration, this identity reduces a six-point integral 
to a sum of six five-point
integrals.  This holds even when $\ell$ is $D$-dimensional.  
The expressions $\omega_0$
and $\omega_j$ are independent of the loop momentum, so for 
purposes of the loop integration
we can treat them as constants, and the division by 
$\omega_0$ introduces no new denominators.

\Eqn{DenominatorIdentity} is not the only equation we can write
down for our integrand;
we could substitute any four vectors for $k_1,\ldots, k_4$ in the 
upper argument to the
Gram determinant.  (Replacing the five momenta in the lower 
argument with any other
set of five external momenta will of course keep the equation 
the same.)  However,
one finds that the new coefficients $\omega'_j$ are related, so 
that their ratios remain
the same up to a term proportional to $G(k_1,k_2,k_3,k_4,k_5)$ --- 
which vanishes.  This
means that we only have one independent relation for the denominators 
in this case.

For larger $n$, we could of course iterate this reduction 
(or its straightforward 
generalization to
integrals with external masses).  Alternatively, we can write down 
all independent
relations directly.  We expect to find $n-5$ independent 
relations.  Modulo accidental
relations, each linear relation should allow us to reduce the 
number of denominators by one.
At the integrand level, without taking into account $\mu^2$ terms 
(which yield integrals
of $\Ord(\eps)$), we expect at best to reduce to terms with five 
denominator factors,
as in the six-point example above.

We can understand the counting intuitively by imagining that we 
have rotated the $D$ dimensional
loop momentum into five dimensions; including any four 
independent external momenta, we expect
to span a space of five independent quantities, which is exactly 
what happens in the above example.

\subsection{Reduction Using Algebraic Geometry}
\label{LoopCAGReductionSection}

In order to analyze real-emission integrands, we will need
to make use of the tools of computational algebraic geometry.
Before doing so, let us first set new lyrics to an old 
melody, recasting the reduction in the previous section into the
language of algebraic geometry.

\def\varset#1{W_{#1}}
In the loop integral case, we take as \textit{variables\/} all 
independent invariants built out of
the loop momentum, a set we shall call $\varset{\ell}$.  
(Momentum conservation gives relations between invariants.)
We take as \textit{parameters\/} all invariants built solely 
out of external
momenta.  We will be interested in polynomial expressions in the 
variables, with coefficients that are
rational functions of the parameters.  

Let us explore the singularities of both sides
of \eqn{DenominatorIdentityII}.  Both sides will have poles when any 
$D_j$ vanishes.  What about singularities were all $D_j$ to vanish 
\textit{simultaneously\/}? In this case,
the right-hand side would have no 
stronger singularity than having only five $D_j$ vanish 
simultaneously. The equation
can only be consistent if the left-hand side, too, has no 
stronger singularity.  This will
be the case
only if all $D_j$ \textit{cannot\/} vanish simultaneously.  At the 
same time, were the external
momenta in arbitrary dimension rather than restricted to four 
dimensions, all denominators
could in fact vanish simultaneously; so this impediment must depend on 
the momenta being
restricted to four dimensions.

The simplest polynomial equation
expressing this restriction is precisely the vanishing of the Gram 
determinant $G_1$.
We thus want to show that the simultaneous equations,
\begin{equation}
D_j = 0\quad (j=1,\ldots,6)\,,\qquad G_1 = 0\,,
\end{equation}
have no solution.  In the language of \textit{ideals}, we want to show that 
the ideal built out of
the $D_j$ and $G_1$ is the unit 
ideal.
  One way to do this, and computationally the most straightforward,
is to compute the Gr\"obner basis of the ideal.  If it is $1$ (or a 
constant), the ideal is
indeed a unit one.  Moreover, there is always a way of expressing the 
elements of the Gr\"obner
basis in terms of the original defining polynomials of the ideal, via 
the \textit{cofactor\/}
matrix.  In the case of a unit ideal, this
takes the following form,
\begin{equation}
c_j D_j + c_0 G_1 = 1\,,
\label{UnityEquation}
\end{equation}
where the $c_i$ are polynomials in the variables $\varset{\ell}$.  
(This is a polynomial analog to B\'ezout's identity.) For the hexagon
integrand,
the coefficients are all constants (rational functions of the 
parameters alone),
\begin{equation}
c_0 = \frac1{\omega_0}\,,\qquad
c_{j\ge 1} = -\frac{\omega_j}{\omega_0}\,.
\end{equation}

In the next section, we review the basics of the computational
tools we will use in later sections to 
partial-fraction real-emission integrands.

\subsection{Algebraic Geometry Review}
\label{ComputationalAlgebraicGeometrySection}

In this section, we provide a brief review 
of elements of computational algebraic geometry needed
for the calculations in this paper. 
Many of these ideas have appeared earlier in the context of loop 
integral reduction, see e.g.{} 
refs.~\cite{Zhang:2012ce,Mastrolia:2012an, Mastrolia:2012wf}.
For a pedagogical introduction to these concepts we refer the 
reader to refs.~\cite{cox1994ideals, cox2013using, Zhang:2016kfo}.

A fundamental object of interest is the ring of polynomials in $n$ variables $x_i$ over a field $\mathbb{F}$, denoted by 
\begin{equation}
   R =  \mathbb{F}[x_1, \ldots, x_n].
\end{equation}
The field $\mathbb{F}$ is often a numeric field, such as the rational or complex numbers, but could also be a field of rational functions. In the context of this paper, the field $\mathbb{F}$ is taken to be the field of rational functions of the observable kinematics. Oftentimes we will work with a fixed, numeric phase-space point, in which case $\mathbb{F}$ reduces again to a numeric field. The tuple of variables $\{x_1, \ldots, x_n\}$ is taken to be the tuple $\{\hat{\lambda}, \tau, s_{i_1 r}, s_{i_2 r}, s_{i_3, r} s_{i_4 r}\}$, where the $i_k$ correspond to distinct partonic or protojet momenta.
Elements of the polynomial ring $R$ are expressed as linear combinations of monomials which we denote as
\begin{equation}
    x^{\vec{\alpha}} = \prod_{i=1}^n x_i^{\alpha_i},
\end{equation}
where the $\alpha_i \in \mathbb{Z}_{\ge 0}$.
It is useful to organize the set of monomials with a monomial ordering, which we denote by $\succ$. This is a total ordering of the exponents $\vec{\alpha}$ of the monomials. Given a monomial ordering $\succ$, any polynomial $p$ has a lead monomial $\mathrm{LM}_{\succ}(p)$, which is either $0$ or largest monomial in $p$ with respect to the ordering $\succ$.
A useful class of orderings are the so-called ``block orderings''. We 
split the variables $x_i$ into two non-overlapping sets 
$\{y_1, \ldots, y_s\}$ and $\{z_1, \ldots z_{n-s}\}$, and to each we 
associate a monomial order $\succ_y$ and $\succ_z$. To compare two 
monomials $x^{\vec{\alpha}}$ and $x^{\vec{\beta}}$, one first compares 
the part of the monomial depending only on $y$ variables using 
$\succ_y$ and in the case of a tie compares the part of the monomials 
depending on the $z$ variables using $\succ_z$.
Unless otherwise specified, we use the degree reverse lexicographic 
ordering for variable blocks in this paper.

\paragraph{Ideal Membership and Gr\"obner Bases}
Given a set of polynomials $\{h_1, \ldots, h_m\}$ in $R$, 
we want to know whether another polynomial $h_0$ can be written as a 
polynomial linear combination of the $h_i$, that is, does there exist a 
set of $c_i \in R$ such that
\begin{equation}
    h_0 = \sum_{i=1}^m c_i h_i.
    \label{eq:IdealMembershipDefinition}
\end{equation}
To answer this question, we introduce the concept of an ideal generated by the $h_i$. This is defined as 
\begin{equation}
    \langle h_1, \ldots, h_m\rangle = \left\{ \sum_{i=1}^m a_i h_i \,\,|\,\, a_i \in R \right\}. 
    \label{eq:idealDefinition}
\end{equation}
This is the set of all polynomial linear combinations of the $h_i$.
We refer to the set $H = \{h_1, \ldots h_m\}$ as generators of the ideal. Given this definition, we can now rephrase the question of the existence of the $c_i$ in \eqn{eq:IdealMembershipDefinition} as ``does 
$h_0$ belong to the ideal $J = \langle h_1, \ldots, h_m \rangle$?''.
To answer this ideal membership question, we introduce the concept of the Gr\"obner basis. Given $J$, a Gröbner basis is a different set of polynomials that also generate $J$. We will denote such a set as
\def\GB{\mathop{\textrm{Gr\"obnerBasis\/}}\nolimits}
\begin{equation}
    \{g_1, \ldots, g_l\}  = \GB{}(H, \{x_1, \ldots, x_n\})
\end{equation}
This generating set allows one to algorithmically check if $h_0 \in J$ by computing the remainder of $h_0$ modulo the Gr\"obner basis. Precisely, given the generating set $H$ and the monomial order $\succ$, it is possible to write any polynomial $p$ as
\begin{equation}
    p = q_1 h_1 + \ldots q_m h_m + \Delta_{H, \succ}(p),
\end{equation}
such that none of the $\mathrm{LM}(h_i)$ factor $\mathrm{LM}(\Delta_{H, \succ}(p))$. That is, the remainder is irreducible by any of the $h_i$. The $q_i$ and $\Delta_{H, \succ}(p)$ can be computed algorithmically by ``polynomial reduction''. Such algorithms are implemented in many common computer algebra systems. 
For brevity, we will say $\Delta_{H, \succ}(p)$ is the remainder of $p$ modulo $H$ and write this as
\begin{equation}
    p = \Delta_{H, \succ}(p) \quad \mathrm{mod} \quad H.
    \label{eq:modNotationDefinition}
\end{equation}
What is special about a Gr\"obner basis
is that having zero remainder modulo such a basis is in one-to-one 
correspondence with ideal membership. That is,
\begin{equation}
    h_0 = 0 \quad \mathrm{mod} \quad \GB{}(H, \{x_1, \ldots, x_n\}) \quad \Leftrightarrow \quad h_0 \in J.
\end{equation}
Moreover, the remainder of a polynomial modulo a Gr\"obner basis is uniquely defined.

\paragraph{Varieties, the Weak Nullstellensatz and the Shape Lemma}

Polynomial ideals are closely related to the surfaces defined by 
setting their generators to zero. 
Specifically, for the ideal $\langle h_1, \ldots, h_m \rangle$ in the ring $R$ we can think of the set 
of elements of $\mathbb{F}^n$ which satisfy the equations $h_i = 0$. To this end, one can define the 
\textsl{variety\/} $V$, 
\begin{equation}
    V(\langle h_1, \ldots, h_m \rangle) = \left\{ a \in \mathbb{F}^n \,\, \text{such that} \,\, h_1(a) = \cdots = h_m(a) = 0 \right\}.
\end{equation}

Hilbert's weak Nullstellensatz says that, if $\mathbb{F}$ is algebraically closed, 
and one can show that the ideal $\langle h_1, \ldots, 
h_m\rangle = \langle 1 \rangle$
(\textit{i.e.\/} it is the unit ideal), then the equations 
\begin{equation}
    h_1 = \cdots = h_m = 0
\end{equation}
have no common zero in $\mathbb{F}^n$. That is, the variety defined by the intersection of the hypersurfaces $h_i = 0$ is empty. 
Importantly, the only (reduced) Gr\"obner basis of $\langle 1 \rangle$ is the set containing the unit polynomial, i.e. $\{1\}$. 
The weak Nullstellensatz therefore provides a geometrical understanding of a Gr\"obner basis computation resulting in the unit ideal.

More generally, we can consider ideals whose associated variety is not empty. It turns out that 
for such an ideal, one can define a geometric dimension which is derived from the 
associated variety. Quite naturally, when the ideal is associated to a finite set of points, 
both the ideal and variety are said to be ``zero-dimensional''. This geometric notion has a 
useful algebraic consequence, when considering remainders modulo zero-dimensional ideals. This is sometimes referred to in the physics literature as the ``shape lemma''~\cite{Mastrolia:2012an}. Let us regard the ring $R$ as an 
infinite-dimensional $\mathbb{F}$-vector space, spanned by monomials in the variables. 
In such an interpretation, if we let $J$ be an ideal and $G$ be a Gr\"obner basis of $J$ with respect to $\succ$,
the operation of computing the polynomial remainder $\Delta_{G, \succ}$ is a linear map acting on $R$. 
For $J$ whose dimension is non-zero, the image of $\Delta_{G, \succ}$ is an infinite-dimensional subspace of $R$.
However, if $J$ is a zero-dimensional ideal, and $\mathbb{F}$ is algebraically
closed, it turns out that the image of $\Delta_{G, \succ}$ is a finite-dimensional subspace of $R$. 
Moreover, the dimension of this vector space is equal to the number of points in the associated variety, 
when counted with appropriate multiplicity.

\paragraph{Co-factors and Syzygies}
For our purposes, it is important to obtain an explicit set of $c_i$ such that the relation \eqn{eq:IdealMembershipDefinition} holds. 
Given a Gr\"obner basis of $J$, and a polynomial $h_0$ in the ideal, polynomial reduction by the Gr\"obner basis explicitly constructs a set of polynomials $b_i$ such that
\begin{equation}
    h_0 = \sum_{i = 1}^l b_i g_i.
\end{equation}
While this is a representation of $h_0$ in terms of generators of the ideal $J$, it is not in terms of the generating set $H$. 
Nevertheless, as the Gr\"obner basis elements belong to the ideal, they are expressible in terms of the generating set $H$.
That is, there exists a set of polynomials $A_{ij}$, such that
\begin{equation}
    g_i = \sum_{j = 1}^m A_{ij} h_j.
    \label{eq:CofactorMatrixDefinition}
\end{equation}
The matrix $A_{ij}$ is known as the co-factor matrix. In the computer algebra system \Singular{}, $A_{ij}$ can be computed by making use of the command \textsf{lift}. 
Together with \eqn{eq:CofactorMatrixDefinition}, this gives us an explicit set of polynomials $c_j$ such that \eqn{eq:IdealMembershipDefinition} holds given by
\begin{equation}
    c_j = \sum_{i = 1}^l b_i A_{ij}.
\end{equation}

A natural observation to make about a relation of the form \eqn{eq:IdealMembershipDefinition} is that the $c_i$ are not uniquely defined. 
Specifically, consider a tuple of polynomials $\{d_1, \ldots , d_m\}$ that satisfies
\begin{equation}
    0 = \sum_{i=1}^m d_i h_i.
    \label{eq:SyzygyDefinition}
\end{equation}
This is known as a {\em syzygy} equation.
Given such a tuple, it is clear that replacing $c_i \rightarrow c_i + d_i$ in \eqn{eq:IdealMembershipDefinition} gives a different representation of $h_0$ in terms of the $h_i$. 
We therefore see that the freedom in the definition of the $c_i$ is parametrized by the 
set of solutions to the \eqn{eq:SyzygyDefinition}. It is clear from the structure of 
\eqn{eq:SyzygyDefinition} that if the tuples $d_i^{(1)}$ and $d_i^{(2)}$ satisfy 
\eqn{eq:SyzygyDefinition}, then so does the tuple $d_i^{(1)}+d_i^{(2)}$. Furthermore, if $x$ 
is any element of $R$ then it is clear that the tuple $x d_i$ satisfies \eqn{eq:SyzygyDefinition}. 
The solutions of \eqn{eq:SyzygyDefinition} have the structure 
of a so-called {\em module\/}. 
It is somewhat useful to consider a module as analogous to a vector 
space, in both cases elements are frequently represented as tuples of elements of a ``scalar'' algebraic 
object. We call this particular 
module the {\em syzygy module\/} of 
$\{h_1, \ldots, h_m\}$.
This syzygy module is a subset of $R^m$ and therefore a submodule of $R^m$. 
Unlike vector spaces, modules do not always have bases. Nevertheless, they do have generating sets and there exist 
algorithms to compute a generating set of a syzygy modules. 
We denote a generating set of the syzygy module of $\{h_1, \ldots, h_m\}$ as $\mathrm{Syz}(h_1, \ldots, h_m)$.

In analogy to Gr\"obner bases in polynomial rings, one can also introduce Gr\"obner bases for modules. 
This will allow us to systematically move through and understand the set of solutions $c_i$ of \eqn{eq:IdealMembershipDefinition}. 
Specifically, we will to be able to find as well as elements with desirable properties such as low degrees.
To this end, let us label the cartesian basis vectors of $R^m$ as $\{e_1, \ldots, e_m\}$. 
Naturally any element of $R^m$ can be expressed linearly in these vectors with elements of $R$ as coefficients. 
One can introduce a monomial ordering on $R^m$ by extending an ordering $\succ$ on $R$.
We will make use of the ``term over position'' extension which we label as $\succ_{\mathrm{TOP}}$. 
We say that $x^{\vec{\alpha}} e_i \succ_{\mathrm{TOP}} x^{\vec{\beta}} e_j$ if $x^{\vec{\alpha}} \succ x^{\vec{\beta}}$ or $x^{\vec{\alpha}}  = x^{\vec{\beta}}$ and $j > i$.
When extending the degree reverse lexicographic ordering, this can be implemented in the \Singular{} Mathematica interface as \{\textsf{DegreeReverseLexicographic}, \textsf{ModuleDescending}\}~\cite{math_singular}.
With a module term ordering, one can once again define a reduction operation, which we will denote in the analogous manner to \eqn{eq:modNotationDefinition}. 
Furthermore, one can also define a Gr\"obner bases with respect to $\succ_{\mathrm{TOP}}$ and
such Gr\"obner bases can be computed in computer algebra systems such as \Singular{}. 
Importantly, if the module ordering extends a degree ordering, then remainders will have the lowest possible degree.

\section{Reduction of Numerator-Free Real-Emission Integrands}
\label{RealEmissionNumeratorFreeReductionSection}

We are ultimately interested in squared real-emission matrix elements
in Yang--Mills theory.  In modern approaches, these would be
computed at the amplitude level using spinor variables, and
then squared analytically or numerically.  We will follow
the same logic as in the general one-loop reduction, and
instead perform a \textit{gedanken\/} calculation of the
integrand.

Each amplitude can be written as a sum of cubic tree diagrams;
we can replace four-point vertices with a connected pair of 
three-point vertices and a numerator factor canceling the
additional denominator.  To study the reduction of numerator-free
integrands, it is then sufficient to study diagrams in $\phi^3$ theory.

Any term in the squared matrix element arises from the product
of two tree diagrams. The analog to a scalar loop integral
is given by the product of two massless cubic diagrams.
Let us begin by considering the
reduction of single diagrams before taking the products that
appear in the squared matrix element. The following discussion is sufficient to show that a finite number of collections of denominators can occur in any real emission integrand.

Given the scheme of dimensional regularization we employ (see \Sect{DimensionalRegularizationSection}), in real-emission contributions, the post-remapping singular
momentum $k_r$ can again be rotated into five 
dimensions; but it has only four independent components, unlike the 
loop momentum, because it is on shell.  The counting of denominators 
is then reduced by one; we expect to reduce an expression with five 
independent denominators to a sum of expressions each with four 
denominators.  We first find five denominators in an eight-point 
amplitude, so that is the analog of the six-point integral that we 
will study.

\begin{figure}
    \includegraphics{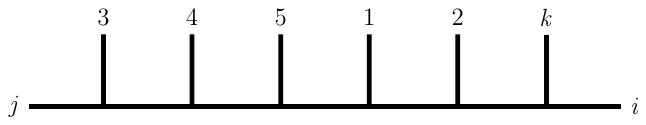}
\caption{An eight-point cubic diagram.
}
\label{EightPointCubic1}
\end{figure}

Each cubic diagram gives rise to one denominator structure.
Many different denominator structures can arise from the ensemble of 
diagrams, of course.  We will classify them and enumerate them
in \Sect{TriskeliaSection}.

The simplest diagram has all three recombining momenta 
adjacent in the ordering of legs.  For example, the diagram shown in 
\fig{EightPointCubic1} gives rise to the following expression,
\begin{equation}
\frac1{s_{j3} s_{j34} s_{j345} s_{j1345} s_{j12345}}\,.
\end{equation}
In this example, we take the momenta $k_1,\ldots,k_5$ to be massless;
the recombining momenta $k_{i,j,k}$ are always massless.

Label the denominators,
\begin{equation}
S_1 = s_{j3}\,,\qquad
S_2 = s_{j34}\,,\qquad
S_3 = s_{j345}\,,\qquad
S_4 = s_{j1345}\,,\qquad
S_5 = s_{j12345}\,.
\end{equation}
For this denominator, the following Gram determinant, 
\begin{equation}
G_2 \equiv G\biggl(\,
\begin{matrix}
k_j,\!&\!k_1,\!&\!\ldots,\!&\!k_{4}\\[-1mm]
k_{5},\!&\!k_1,\!&\!\ldots,\!&\!k_{4}
\end{matrix}
\biggr) 
\end{equation}
\def\realco{\omega^r}
gives rise to an equation analogous
to \eqn{DenominatorIdentity},
\begin{equation}
\frac{G_2}{S_1 S_2 S_3 S_4 S_5} = 0 = 
\sum_{j=1}^5 \frac{\realco_j}{S_1\cdots \Xed{S}_j\cdots S_5}
+\frac{\realco_0}{S_1 S_2 S_3 S_4 S_5}\,, 
\label{PartialFractionI}
\end{equation}
where,
\begin{equation}
\begin{aligned}
\realco_1 &= G\biggl(\,
\begin{matrix}
k_1,\!&\!k_2,&\!k_3,\!&\!k_{4}\\[-1mm]
k_1,\!&\!k_2,&\!k_3+k_4,\!&\!k_{5}
\end{matrix}
\biggr)\,,
\\
\realco_2 &= -G\biggl(\,
\begin{matrix}
k_1,\!&\!k_2,&\!k_3,\!&\!k_{4}\\[-1mm]
k_1,\!&\!k_2,&\!k_3,\!&\!k_{4}+k_5
\end{matrix}
\biggr)\,,
\\
\realco_3 &= -G\biggl(\,
\begin{matrix}
k_1,\!&\!k_2,&\!k_3,\!&\!k_{4}\\[-1mm]
k_2,\!&\!k_3,&\!k_4,\!&\!k_1+k_{5}
\end{matrix}
\biggr)\,,
\\
\realco_4 &= G\biggl(\,
\begin{matrix}
k_1,\!&\!k_2,&\!k_3,\!&\!k_{4}\\[-1mm]
k_1+k_2,\!&\!k_3,&\!k_4,\!&\!k_{5}
\end{matrix}
\biggr)\,,
\\
\realco_5 &= -G\biggl(\,
\begin{matrix}
k_1,\!&\!k_2,&\!k_3,\!&\!k_{4}\\[-1mm]
k_1,&\!k_3,&\!k_4,\!&\!k_{5}
\end{matrix}
\biggr)\,;
\end{aligned}
\end{equation}
and,
\begin{equation}
\realco_0 = -s_{45}\, \realco_3 -s_{145}\, \realco_4 -s_{1245}\, \realco_5\,.
\end{equation}

While $G_2$ vanishes on physical configurations because the
momenta $k_1,\ldots, k_5$ are four-dimensional, it does not
vanish manifestly when expressed in terms of invariants.  In this
way, it gives rise to a partial-fractioning identity in
\eqn{PartialFractionI}.

Other diagrams, however, give rise to denominators where no Gram 
determinant (nor even a set of Gram determinants involving the partons 
alone) suffices to obtain a partial-fractioning identity.  
For such diagrams, we need to deploy the machinery of
computational algebraic geometry reviewed in
\sect{ComputationalAlgebraicGeometrySection}.

\subsection{Reduction Using Algebraic Geometry}

\begin{figure}
    \includegraphics{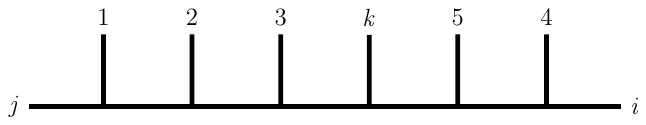}
\caption{A second eight-point cubic diagram.
}
\label{EightPointCubic2}
\end{figure}

Let us now consider the general case of real emission integrand partial fractioning, 
where we cannot simply rearrange a Gram-determinant identity. An example is the term,
\begin{equation}
\frac1{s_{i4} s_{i45} s_{j1} s_{j12} s_{j123}}\,,
\label{Denominator2}
\end{equation}
which arises from the diagram in \fig{EightPointCubic2}.

In the abstract,
with $T_1=s_{i4}$, $T_2=s_{j1}$, $T_3=s_{i45}$, 
$T_4=s_{j12}$, and $T_5=s_{j123}$,
we seek an identity of the form,
\begin{equation}
0 = 
\frac{c_0}{T_1 T_2 T_3 T_4 T_5}\, 
+\sum_{j=1}^5 \frac{c_j}{T_1\cdots \Xed{T}_j\cdots T_5}
+\sum_{j=1}^{n_{z}} \frac{\hat c_j Z_j}{T_1 T_2 T_3 T_4 T_5}\,,
\label{eq:RealEmissionReductionExample}
\end{equation}
where the $Z_j$ are functions of the variables which vanish
on physical configurations.  These functions would implement
the four-dimensionality of the protojets or the dependence of
the protojets on the recombining momenta.  As variables,
we take a set of invariants dependent on the
recombining momenta including the $T_i$, along with 
the mapping coefficients $c_{\ah,i}$, $c_{\ah,k}$, and $\lambda$
appearing in \eqn{CoefficientEquationPair}.  Because we impose
the four-dimensionality via Gram constraints, we need 20
invariants, the number independent in $D$ dimensions, rather
than the 14 independent in four dimensions.

More concretely, we seek an identity of the form,
\begin{equation}
0 = 
\frac{c_0}{T_1 T_2 T_3 T_4 T_5}\,
+\sum_{j=1}^5 \frac{c_j}{T_1\cdots \Xed{T}_j\cdots T_5}
+\sum_{j=3}^6 \frac{\hat c_j G_j}{T_1 T_2 T_3 T_4 T_5}
+\sum_{j=1}^4 \frac{\breve c_j R_j}{T_1 T_2 T_3 T_4 T_5}
\end{equation}
where the Gram determinants $G_{3}, \ldots, G_{6}$ are,
\begin{equation}
\begin{aligned}
G_3 &\equiv G\biggl(\,
\begin{matrix}
k_{\ah},\!&\!k_1,\!&\!\ldots,\!&\!k_{4}\\[-1mm]
k_{\ah},\!&\!k_1,\!&\!\ldots,\!&\!k_{4}
\end{matrix}
\biggr)\,,
\\ G_4 &\equiv G\biggl(\,
\begin{matrix}
k_{\bh},\!&\!k_1,\!&\!\ldots,\!&\!k_{4}\\[-1mm]
k_{\bh},\!&\!k_1,\!&\!\ldots,\!&\!k_{4}
\end{matrix}
\biggr)\,,
\\ G_5 &\equiv G\biggl(\,
\begin{matrix}
k_i,\!&\!k_1,\!&\!\ldots,\!&\!k_{4}\\[-1mm]
k_{5},\!&\!k_1,\!&\!\ldots,\!&\!k_{4}
\end{matrix}
\biggr)\,,
\\ G_6 &\equiv G\biggl(\,
\begin{matrix}
k_k,\!&\!k_1,\!&\!\ldots,\!&\!k_{4}\\[-1mm]
k_{5},\!&\!k_1,\!&\!\ldots,\!&\!k_{4}
\end{matrix}
\biggr)\,,
\end{aligned}
\end{equation}
and the remapping-constraint functions $R_i$ are given
by \eqns{RemappingConstraint}{RemappingConstraint4D}.  
The Grams $G_{3,4}$ enforce the four-dimensionality of
$k_{\ah,\bh}$, and $G_{5,6}$ remove otherwise vanishing
terms from the $c_j$.  The remapping constraints $R_{1,2}$
enforce the on-shell conditions for $k_{\ah,\bh}$, and $R_{3,4}$
implement the appropriate variant of dimensional regularization.
Unlike the case of \eqn{PartialFractionI},
we cannot obtain such a decomposition if 
we insist that the $c_j$ be functions of the nonrecombining
invariants alone.  We would however expect to
obtain an identity of the form of 
\eqn{eq:RealEmissionReductionExample} if
we allow the $c_j$ to be polynomials in the variables.

In order to find such a decomposition, we would need to
show that the joint equations,
\begin{equation}
T_j = 0\quad (j=1,\ldots,5)\,,\qquad G_j = 0\quad (j=3,\ldots,6)\,,
\qquad R_j=0\quad (j=1,\ldots,4)\,,
\end{equation}
have no solution.  With presently available Gr\"obner-basis
algorithms and implementations, this problem appears 
hopelessly intractable
if attempted in a fully analytic way.  We use the computational
algebra system \Singular{}, but we expect the performance
of other available systems to be similar.

In order to accomplish our goal, we must simplify the 
calculation.  We do so as follows: we switch to variables
built entirely out of $k_{\ah}$, $k_{\bh}$, and $k_r$ in
addition to the partons $k_{1,\ldots5}$; 
we solve the Gram-determinant constraints explicitly;
we use finite-field
numerical values for all momenta except $k_r$; and we use modular
arithmetic in \Singular{}, which yields vastly faster
Gr\"obner-basis computation than with rational arithmetic.

In this approach, we should think of 
the recombining momenta $k_i, k_j, k_k$ purely
as functions of $k_{\ah}$, $k_{\bh}$, and $k_r$ via the 
inverse mapping~\eqn{InverseAntennaMapping1}.  
The singular momentum $k_r$ is the
only variable momentum; all others are held fixed to specific
finite-field values.  
That is, we do \textit{not\/} choose values for the
recombining partons $k_i$, $k_j$, and $k_k$; nor do we choose
a specific form for the $\lamh$ function in the inverse mapping.
We will take as our variables a set of independent
invariants built out of the $D$-dimensional momentum $k_r$,
along with the symbols $\tau$ and $\lamh$.  These take
the place of the set $\varset{\ell}$ in the one-loop
integrand reduction.

Choosing numerical values for $k_{1,\ldots,5}$ and $k_{\ah,\bh}$
satisfies the constraints $G_3 = 0 = G_4$ automatically; we can
solve the constraint $G_6=0$ by choosing a set of four
invariants, for example 
$\varset{r}\equiv\{s_{\ah r}, s_{r\bh}, s_{r1}, s_{r2}\}$,
and expressing the other invariants involving $r$ in terms of them,
\begin{equation}
    s_{rc} = 
    \frac{G\Bigl(\gatop{c,\bh,1,2}{\ah,\bh,1,2}\Bigr)}
                  {G({\ah,\bh,1,2})} s_{\ah r}
    +\frac{G\Bigl(\gatop{\ah,c,1,2}{\ah,\bh,1,2}\Bigr)}
                  {G({\ah,\bh,1,2})} s_{r\bh}
    +\frac{G\Bigl(\gatop{\ah,\bh,c,2}{\ah,\bh,1,2}\Bigr)}
                  {G({\ah,\bh,1,2})} s_{r1}
    +\frac{G\Bigl(\gatop{\ah,\bh,1,c}{\ah,\bh,1,2}\Bigr)}
                  {G({\ah,\bh,1,2})} s_{r2}\,,
                  \qquad (c=3,4,5)\,.
\end{equation}
As we have chosen numerical values for $k_{1,\ldots 5}$,
the coefficients are just numbers.
(Because $k_r$ is $D$-dimensional, there is no further relation
between these invariants arising from $k_r^2=0$.)
The vanishing of $G_5$ then follows as well.

The several $R_j$ are replaced by the lone constraint 
$\Rtau = 0$, with $\Rtau$ defined in \eqn{TauConstraint}.
In addition, we must supply a polynomial constraint equation
fixing the functional form of $\lamh$; a convenient choice is,
\def\Rlamh{R_{\lamh}}
\begin{equation}
\Rlamh \equiv \lamh (s_{r\ah}+s_{r\bh})+s_{r\bh}-s_{r\ah}\,,
\end{equation}
with $\lamh$ defined by $\Rlamh = 0$.  Other choices are possible,
but we find that this one gives the simplest identities.
It also simplifies the Jacobian~\eqref{InverseAntennaJacobian}.
As variables for computing the Gr\"obner basis, we take the four
variables in $\varset{r}$, along with $\tau$ and $\lamh$, 
\begin{equation}
   V = \varset{r} \cup \{\tau,\lamh\}
     = \{s_{r\ah}, s_{r\bh}, s_{r1}, s_{r2}, \tau, \lamh\} \,. 
\label{RealEmissionVariableSet}
\end{equation}
Our ideal is then $\langle B\rangle$, where,
\begin{equation}
B = \{T_1,T_2,T_3,T_4,T_5,\Rtau,\Rlamh\}\,.
\label{IdealBasisCubic2}
\end{equation}
One might wonder if this setup is $D$-dimensional, as we only include variables with a four-dimensional dependence on $k_r$. However, we also do not include the on-shell condition for $k_r$ in $B$. This is consistent as, due to the spherical symmetry beyond 4-dimensions, the $\epsilon$ dimensional components of $k_r$ can be used to explicitly solve the on-shell condition. Hence the setup is indeed $D$-dimensional.

Recalling the discussion of \sect{ComputationalAlgebraicGeometrySection}, we see that the first step to finding a partial fractions identity is to compute the Gr\"obner basis,
\begin{equation}
    \GB{}\bigl( B;V\bigr)\,;
\end{equation}
in \Singular{}, we use the \textsf{slimgb\/} function for
$\GB{}$.
We use a block ordering with the four invariants forming one
block, and $\{\tau,\lamh\}$ the other, 
and reverse degree lexicographic (degrevlex) within each block.

\def\lindent{\hspace*{3mm}}
We find that the Gr\"obner basis is indeed $\{1\}$, which
shows that the desired equations have no common solution, and
that a partial-fractioning identity then exists. To find the desired identity, we make use of the fact that the cofactor matrix relates the original generating set to the Gr\"obner basis.
We can obtain the cofactor matrix $C$ using \Singular{}'s 
\textsf{lift\/} function.
By definition,
\begin{equation}
    C\cdot B \equiv \sum_i C_i B_i = 1\,.
\end{equation}
This offers us an identity of the form,
\begin{equation}
    \frac1{T_1 T_2 T_3 T_4 T_5} =
\sum_{j=1}^5 \frac{C_j}{T_1\cdots \Xed{T}_j\cdots T_5}
+\frac{C_6 \Rtau}{T_1 T_2 T_3 T_4 T_5}
+\frac{C_7 \Rlamh}{T_1 T_2 T_3 T_4 T_5}\,.
\label{Cubic2Identity1}
\end{equation}
This is of the desired form,
and already allows us to reduce
terms with five-factor denominators to a sum of terms
with four-factor denominators (as the $\Rtau$ and $\Rlamh$
terms will ultimately vanish on physical configurations).

Nevertheless, this identity is not optimal as the $C_j$
(for $j=1,\ldots,5$)
are not necessarily independent of the variables in $\varset{r}$.
This risks introducing powers
of $k_r$ into the numerators.  
To address this issue, we recall from \sect{ComputationalAlgebraicGeometrySection} that $C$ is in fact only
well-defined up to \textit{syzygies\/} of $B$.  If $z$ is a
syzygy of $B$, that is if $z\cdot B = 0$, then,
\begin{equation}
    (C+z)\cdot B = C\cdot B = 1\,.
    \label{SyzygyShift}
\end{equation}
It is therefore clear that we can shift $C$ to reduce or eliminate powers
of $k_r$ in its first five components. To do this systematically, we 
again make use of Gr\"obner basis techniques. Specifically, the 
remainder of polynomial division by a degree ordered Gr\"obner basis, 
is always of the lowest possible degree. Therefore,   
in order to obtain
the ``simplest'' form for $C$, we will reduce it against a modified
Gr\"obner basis of the syzygies of $B$.  The remainder will be
the desired simplest form.

\def\Syz{\mathop{\textrm{Syz}}\nolimits}

We first compute the syzygy module of $B$, which we can do in 
\Singular{} using the \textsf{syz\/} function,
\begin{equation}
    \Syz(B) = \textsf{syz\/}(B)\,.
\end{equation}
We now wish to use this syzygy module in a Gr\"obner basis computation to find an alternative choice for $C$, with lower polynomial degree. 
A subtlety in implementing this is that we are uninterested in the values of $C_6$ and $C_7$ as they 
multiply the constraint functions $R_\tau$ and $R_\lamh$ which are zero on physical configurations. 
Nevertheless, we find that the value of $\{C_1, \ldots, C_5\}$ with the lowest polynomial degree 
does not come together with the $\{C_6, C_7\}$ with the lowest polynomial degree. For this reason, 
we are forced to instruct the Gr\"obner basis algorithm to throw away $C_6$ and $C_7$.  
To this end, we modify the syzygy module by adding two syzygies which freely change the last two 
components of $B$, corresponding to $\Rtau$ and $\Rlamh$,
\begin{equation}
\Syz'(B) = \Syz(B) \cup \{ (0,0,0,0,0,1,0), (0,0,0,0,0,0,1)\}\,.    
\label{eq:SyzModuleModification}
\end{equation}
In any degree ordering, this has the effect that the components $C_6$ and $C_7$ will always reduce to zero, and have therefore been discarded within the computation.
This procedure could be phrased more elegantly and formally in terms
of quotient rings.

\def\Cred{{\overline {\kern -1.5pt C\kern 1pt}}}
\def\GBSyz{\mathop{\textrm{GBSyz}}\nolimits}
\def\rem{\mathop{\textrm{rem}}\nolimits}
We now compute the Gr\"obner basis of the modified syzygy module,
\begin{equation}
    \GBSyz{}'(B) = \GB{}(\Syz{}'(B); V)\,.
\end{equation}
In computing this Gr\"obner basis, we make use of a term over position 
module ordering which extends a block ordering with the four invariants forming one
block, and $\{\tau,\lamh\}$ the other. Within each block, we use reverse degree 
lexicographic (degrevlex). The block nature of the ordering means that the reduction
procedure first tries to minimize the degree of the invariants, and once this is 
minimal tries to minimize the degree of $\tau$ and $\lamh$. We then reduce $C$ against this basis,
\begin{equation}
    \Cred = C\!\,\mod\, \GBSyz'(B)
\end{equation}
That is, we repeatedly divide by elements of the Gr\"obner basis,
finding the remainder, until the operation terminates.  Because
of the nature of a Gr\"obner basis, this procedure terminates
and gives a unique answer in spite of the subtleties of
multivariate division.  We do the whole reduction in one
step using \Singular{}'s
\textsf{reduce\/} function.  For the basis $B$ given
in \eqn{IdealBasisCubic2}, the denominator factors are,
\begin{equation}
\begin{aligned}
T_1 &= s_{\ah r} \left(\tau  \left(\lamh n_{18}+n_{17}\right) s_{r1}
  +\tau  \left(\lamh n_{24}+n_{23}\right) s_{r2}
  +\tau  \left(\lamh n_{12}+n_{11}\right) s_{r\bh}
  +\lamh n_5 \tau +n_4 \tau +n_3\right)
\\&\hphantom{=}
  +\tau  s_{\ah r}^2 \left(\lamh n_7+n_6\right)
  +s_{r\bh} \left(\tau  \left(\lamh n_{20}+n_{19}\right) s_{r1}
  +\tau  \left(\lamh n_{26}+n_{25}\right) s_{r2}+\lamh n_{10} \tau 
  +n_9 \tau +n_8\right)
\\&\hphantom{=}
  +n_{16}\lamh  \tau  s_{r1}+s_{r2} 
  \left(\lamh n_{22} \tau +n_{21}\right)
  +\tau  \left(\lamh n_{14}+n_{13}\right) s_{r\bh}^2
  +n_{15} s_{r1}+n_2 \tau +n_1\,,\\
T_2 &= s_{r1}\,,\\
T_3 &= s_{\ah r} \left(\tau  \left(\lamh n_{44}+n_{43}\right) s_{r1}
+\tau  \left(\lamh n_{50}+n_{49}\right) s_{r2}
+\tau  \left(\lamh n_{38}+n_{37}\right) s_{r\bh}
+\lamh n_{31} \tau 
\right. \\&\hphantom{=}\left. 
    \hphantom{s_{\ah r} ()}\vphantom{\tau\lamh n_1}
+n_{30} \tau +n_{29}\right)
+n_{42}\lamh  \tau  s_{r1}
+s_{r2} \left(\lamh n_{48} \tau +n_{47}\right)
\\&\hphantom{=}
+\tau  s_{\ah r}^2 \left(\lamh n_{33}+n_{32}\right)
+s_{r\bh} \left(\tau  \left(\lamh n_{46}+n_{45}\right) s_{r1}
+\tau  \left(\lamh n_{52}+n_{51}\right) s_{r2}+\lamh n_{36} \tau 
\right. \\&\hphantom{=}\left. 
    \hphantom{s_{\ah r} ()}\vphantom{\tau\lamh n_1}
+n_{35} \tau +n_{34}\right)
+\tau  \left(\lamh n_{40}+n_{39}\right) s_{r\bh}^2+n_{41} s_{r1}
+n_{28} \tau +n_{27} \,,\\
T_4 &= n_{53}+s_{r1}+s_{r2}\,,\\
T_5 &= n_{55} s_{\ah r}+n_{57} s_{r1}+n_{58} s_{r2}
+n_{56} s_{r\bh}+n_{54}\,,\\
\end{aligned}    
\end{equation}
and the first five
components of the reduced cofactor [one-dimensional] matrix after this
procedure are,
\def\Cb{\overline{C}}
\begin{equation}
\begin{aligned}
\Cb_1 &= \lamh \left(n_{60}-\tau  \left(n_{64} \tau +n_{62}\right)\right)-\tau \left(n_{63} \tau +n_{61}\right)+n_{59}\,,\\
\Cb_2 &= \lamh^2 \tau  \left(n_{72} \tau +n_{69}\right)
+\lamh \left(\tau  \left(\tau  \left(n_{73} \tau   
+n_{71}\right)+n_{68}\right)+n_{66}\right)
+\tau  \left(n_{70} \tau +n_{67}\right)-n_{65}\,,\\
\Cb_3 &= -\lamh \left(\tau  \left(n_{77}-n_{79} \tau\right)
+n_{75}\right)-\tau \left(n_{78} \tau +n_{76}\right)-n_{74}\,,\\
\Cb_4 &= \lamh^2 \tau  \left(n_{84}
   -\tau \left(n_{89} \tau+n_{87}\right)\right)
+\lamh \left(n_{81}-\tau  \left(\tau  \left(n_{88} \tau +n_{86}\right)+n_{83}\right)\right)
-\tau \left(n_{85} \tau +n_{82}\right)+n_{80}\hspace*{-10mm}\,,\\
\Cb_5 &= \lamh^2 \tau \left(n_{94}-\tau  \left(n_{99} \tau 
+n_{97}\right)\right)+\lamh \left(\tau  \left(n_{93}
-\tau  \left(n_{98} \tau +n_{96}\right)\right)
+n_{91}\right)+\tau  \left(n_{95} \tau +n_{92}\right)+n_{90}\,,
  \hspace*{-10mm}\\
\end{aligned}
\end{equation}
where the $n_i$ are numerical (finite-field)
values.  Values for a particular
momentum configuration are given in
appendix~\ref{SampleConfigurationAppendix}.
As we modified the syzygy module in \eqn{eq:SyzModuleModification}, $\overline{C}_6$ and $\overline{C}_7$ are trivially zero. 
In any event, they multiply functions which are zero on physical configurations and are so of no interest.
Interestingly, the reduction removes dependence on the variables in $\varset{r}$
from $C_{1,\ldots,5}$; but it does not remove dependence on
$\tau$ and $\lamh$. Note that this is a consequence of the chosen monomial ordering and choice of $\hat{\lambda}$. 
We will see in \sect{ResultsSection} that this is a general feature of the partial fractions approach.

\section{Triskelia}
\label{TriskeliaSection}

In the previous section, we discussed an algorithm to construct
partial-fraction identities for a 
given denominator structure.  We turn now to the task of systematically 
organizing the set of denominators
that arise in a tree-level amplitude.  We focus on an organization
that will allow us to characterize all different ways a denominator
can depend on the singular momentum when using
the inverse antenna mapping~\eqref{InverseAntennaMapping1} to
express the recombining momenta in terms of the protojet and
singular momenta.
This will then allow to systematically collect partial-fraction
identities for all possible denominators in a tree-level amplitude.
This dependence 
can be graphically organized by combinatoric devices that we dub \textsl{triskelia\/}.  The best-known triskelion in the
wider culture is probably the Isle of Man's.

Our starting point is to consider the complete set of tree-level 
Feynman diagrams for an $n$-point amplitude in a cubic theory.
As discussed earlier, this suffices for Yang--Mills.  
The external legs are the recombining momenta
$k_{i,j,k}$ as well as $n-3$ numbered momenta.  Using momentum
conservation, each propagator denominator can be rearranged
to depend on at most one of the three recombining momenta.  The
denominator factors independent of them are simply constants,
and we can pull them out front as overall prefactors.

\begin{figure}[p]
    \centering
    \begin{tabular}{c@{\hspace{4mm}}c@{\hspace{4mm}}c}
    \includegraphics[width=0.3\textwidth]{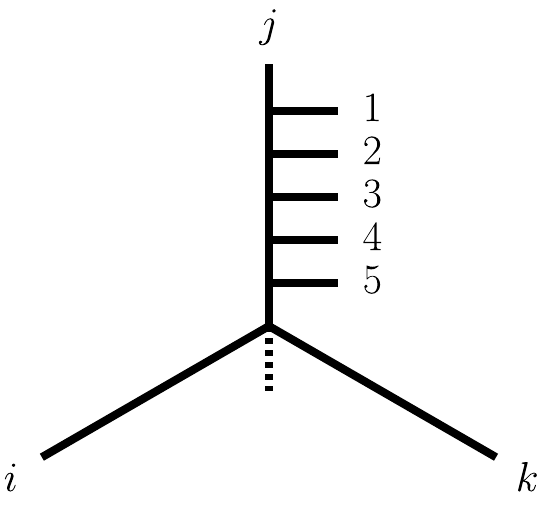}& 
    \includegraphics[width=0.3\textwidth]{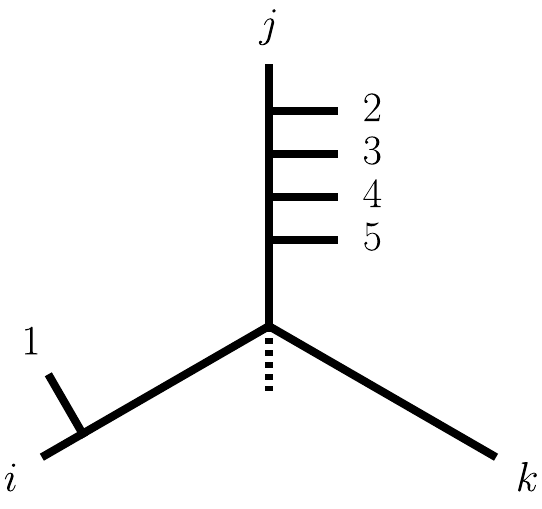} &
    \includegraphics[width=0.3\textwidth]{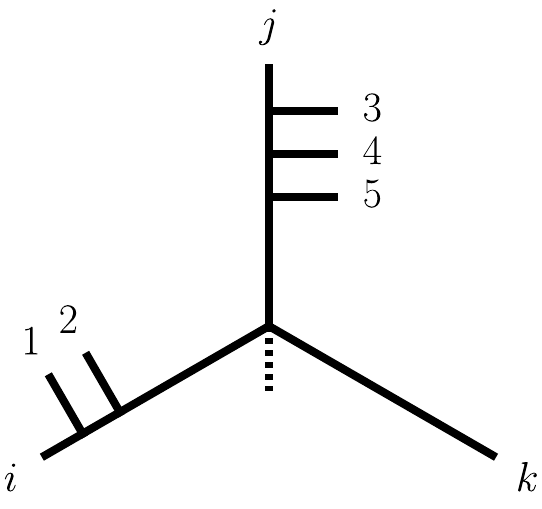}
    \\[-4mm] (a) & (b) & (c)
    \\[2mm]
    {\includegraphics[width=0.3\textwidth]{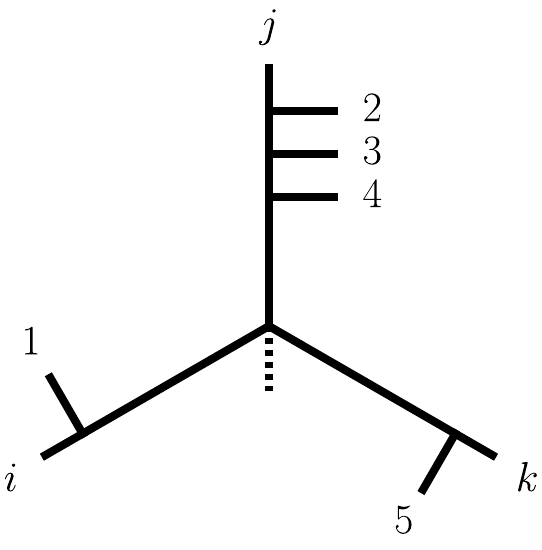}} &
    \raisebox{3mm}{\includegraphics[width=0.3\textwidth]{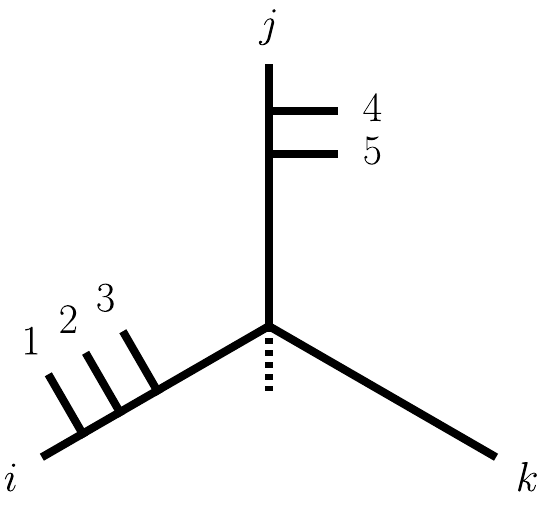}} &
    {\includegraphics[width=0.3\textwidth]{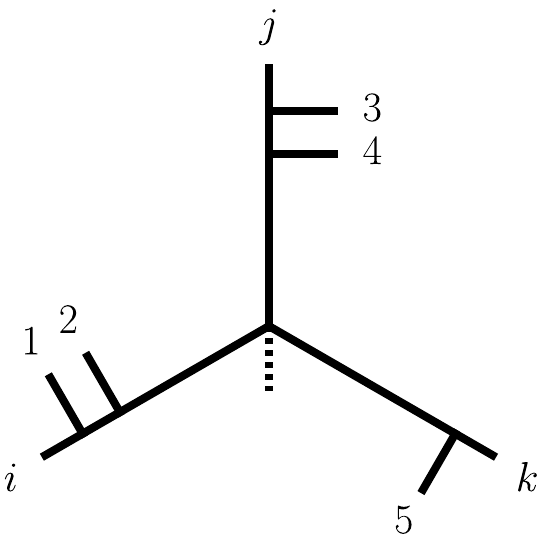}}
    \\[-4mm] (d) & (e) & (f)
    \\[2mm]
    \raisebox{3mm}{\includegraphics[width=0.3\textwidth]{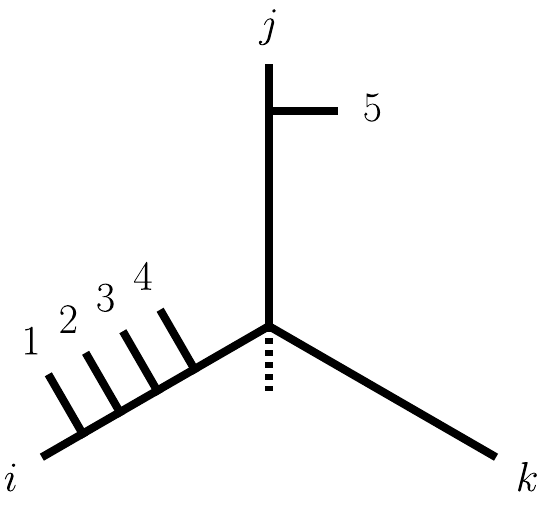}} &
    \includegraphics[width=0.3\textwidth]{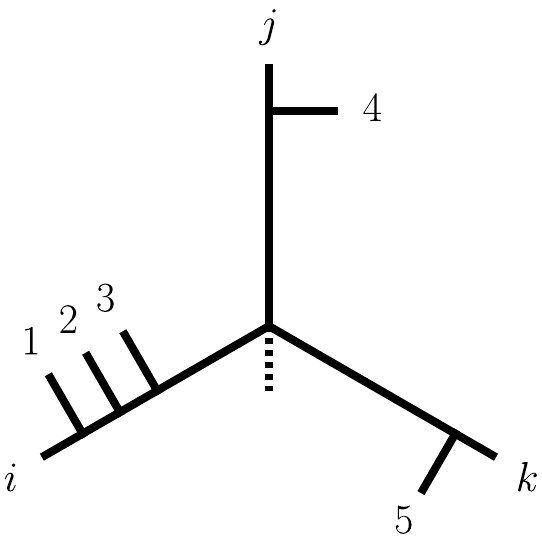} &
    \includegraphics[width=0.3\textwidth]{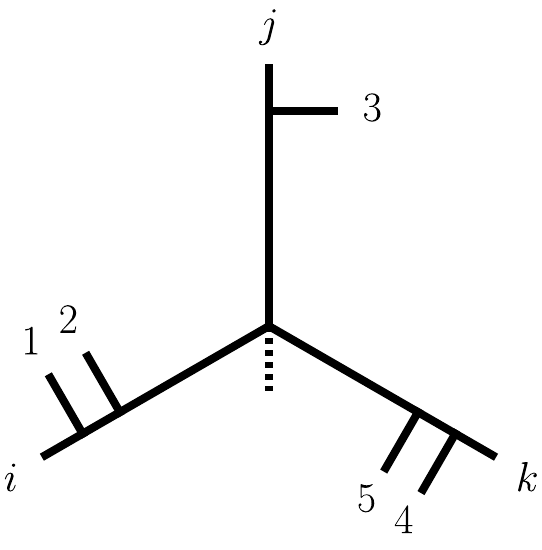}
    \\[-4mm] (g) & (h) & (i)
    \\[2mm]
    \raisebox{3mm}{\includegraphics[width=0.3\textwidth]{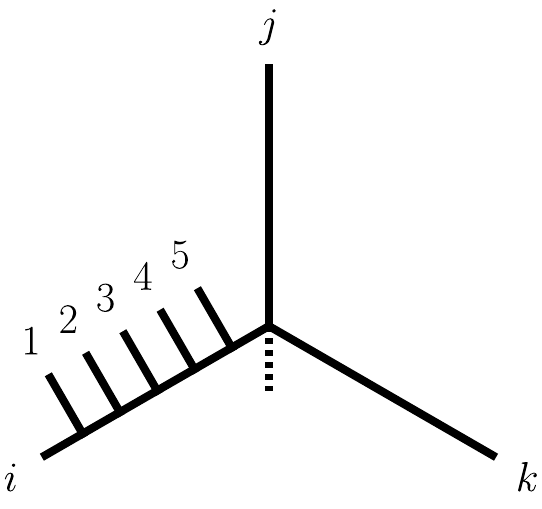}} &
    \includegraphics[width=0.3\textwidth]{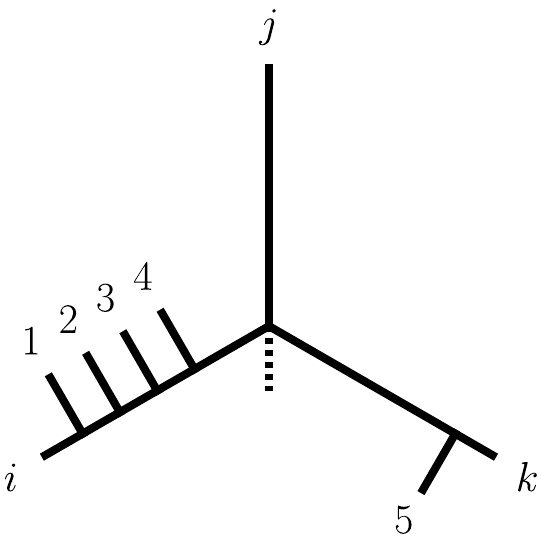} &
    \includegraphics[width=0.3\textwidth]{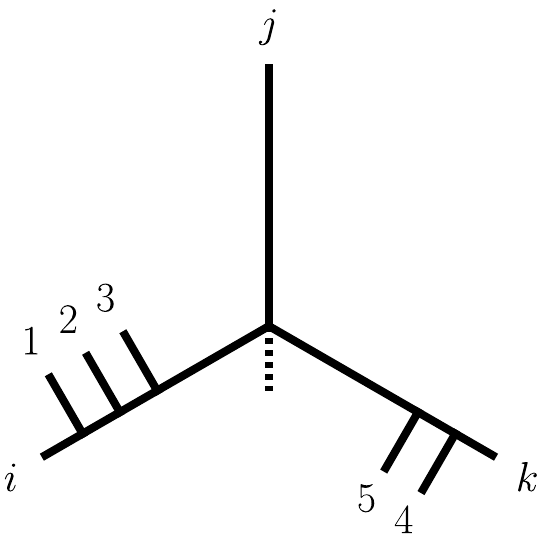}
    \\[-4mm] (j) & (k) & (l)
    \\[2mm]
    \end{tabular}
    \caption{Triskelia with five denominators: (a) $\triskelion500$
    (b) $\triskelion410$ (c) $\triskelion320$ (d) $\triskelion311$
    (e) $\triskelion230$ (f) $\triskelion221$ (g) $\triskelion140$
    (h) $\triskelion131$ (i) $\triskelion122$ (j) $\triskelion050$
    (k) $\triskelion041$ (l) $\triskelion032$. Solid external
    lines are either massless or massive, while internal lines are
    all massless.  The dotted line represents either no external
    leg, or either a massless or massive external leg.} 
    \label{fig:fiveDenominatorTriskelia}
\end{figure}

We can
re-express the recombining momenta in terms of the protojet
momenta $k_{\ah,\bh}$ and the singular momentum $k_r$ using
the inverse antenna mapping~\eqref{InverseAntennaMapping1}.  At
this point, each denominator factor dependent on a recombining momentum
depends implicitly on the singular momentum.  (The denominator
factors independent of the recombining momenta are also independent
of the singular momentum.)
In any given Feynman diagram, the 
denominator factors dependent on the singular momentum
can therefore be organized 
into three groups: one for each recombining momentum. Analogously to similar practices when considering 
loop diagrams, we will draw only the propagators corresponding to these 
denominators, ``pinching'' the others. In any such pinched diagram, 
each of the three groups forms a line, and the three lines meet at a 
single ``central vertex''. 
We can thus visualize a denominator structure in terms of a 
three-pronged object, that is a \textsl{triskelion\/}. 
Examples of such triskelia can be found in
\figs{fig:fiveDenominatorTriskelia}{fig:fourDenominatorTriskelia}.

In the following sections, we will investigate the 
partial-fraction decompositions by 
enumerating the identities for all distinct
triskelia with a fixed number of denominators.  
As an aid in this endeavor, 
let us count the number of kinematically distinct triskelia with $n$ 
denominators. Each of 
these denominators must belong to one of the $k_i$, $k_j$ or $k_k$ 
denominator groups. 
We begin by counting the number of ways to partition the $n$ 
denominators into three groups. 
By a standard ``stars and bars'' argument\footnote{The 
partitioning of $n$ 
objects into $d$ groups is the same counting problem as finding all 
monomials of degree $n$ in $d$ variables, hence the counts agree.}, 
it is easy to show that for $d$ groups 
this is $\binom{n+d-1}{d-1}$. There are therefore $\binom{n+2}{2}$ 
possible ``skeleton'' 
triskelia. We next consider these as skeletons, 
as we have not yet discussed the configurations of 
momenta flowing into the diagram.   
For each of the corresponding denominator structures, 
there can be either (a) no momentum, (b) a massless momentum, or 
(c) a massive momentum flowing 
into the central vertex of the triskelion. 
Furthermore, each of the $n-3$ numbered momenta can be
either massless or massive. We therefore find that
\begin{equation}
    \text{\# triskelia with } n \text{ denominators} = 3 \times \binom{n+2}{2} \times 2^n.
    \label{eq:TriskeliaCount}
\end{equation}

We denote a triskelion by the notation $\triskelion{n_j}{n_i}{n_k}$,
where $n_j$ is the number of external momenta attached directly
to the same central-vertex line as $k_j$, and similarly $n_{i,k}$
count the number of external momenta attached to the same
lines as $k_{i,k}$ respectively.  This notation does not
distinguish whether the external momenta are massive or massless,
nor the number or type attached directly to the central vertex.
We can reduce the number of triskelia to be considered by 
relying on the $n_i\leftrightarrow n_k$ symmetry, and choosing
$n_i\ge n_k$.  For $n=5$, 
there are 1152 triskelia that we must consider.

\def\tsp{\hspace*{4mm}}

\begin{figure}[th]
    \centering
    \begin{tabular}{c@{\hspace{4mm}}c@{\hspace{4mm}}c}
    \includegraphics[width=0.3\textwidth]{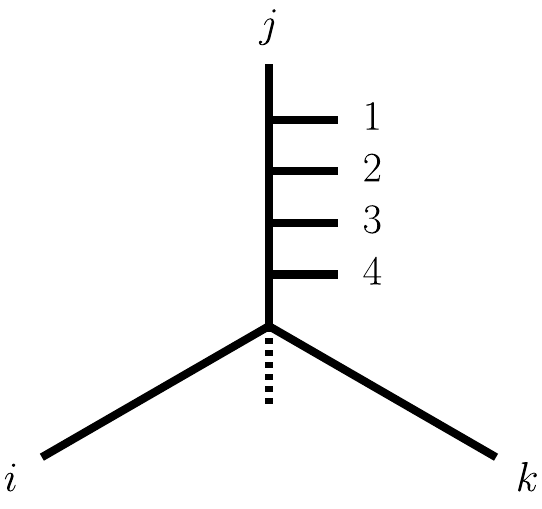} &
    \includegraphics[width=0.3\textwidth]{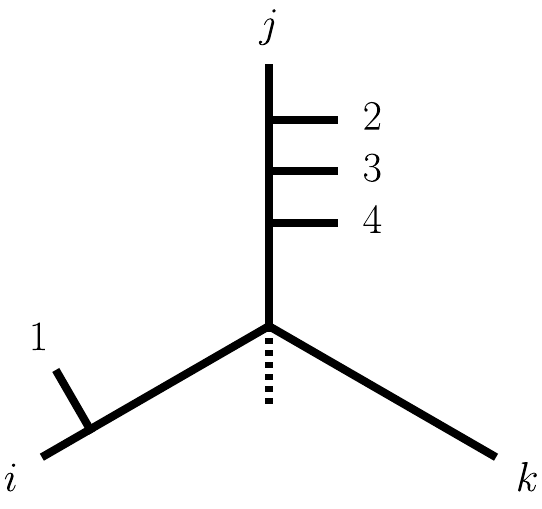} &
    \includegraphics[width=0.3\textwidth]{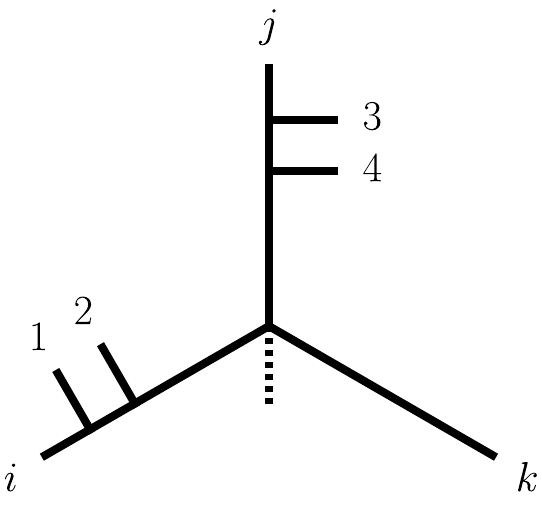}
    \\[-4mm] (a) & (b) & (c)
    \\ 
    \includegraphics[width=0.3\textwidth]{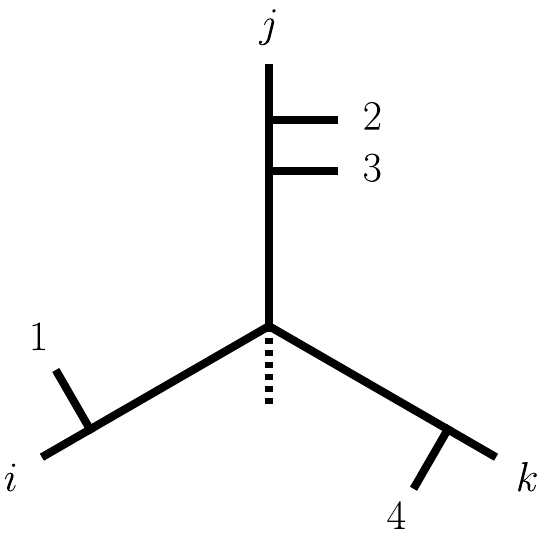} &
    \includegraphics[width=0.3\textwidth]{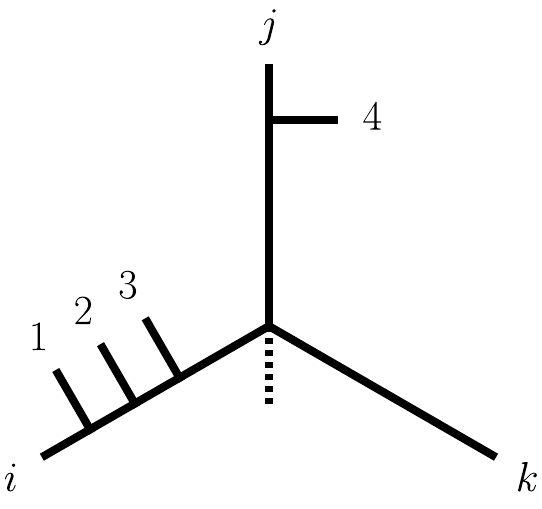} &
    \includegraphics[width=0.3\textwidth]{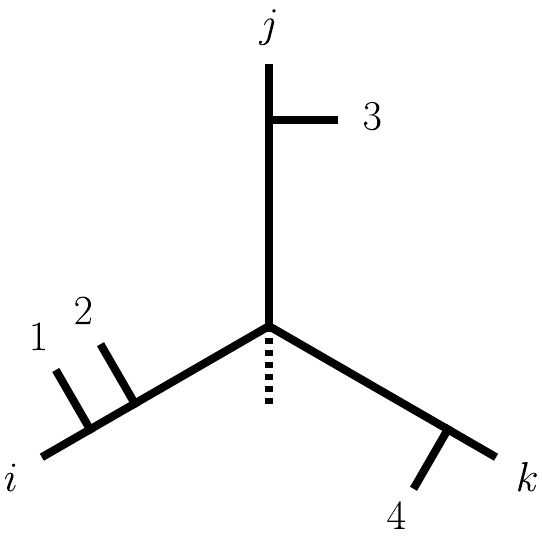}
    \\[-4mm] (d) & (e) & (f)
    \\ 
    \includegraphics[width=0.3\textwidth]{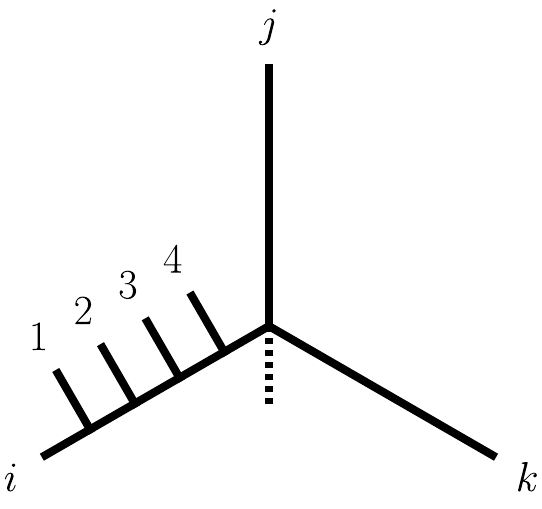} &
    \includegraphics[width=0.3\textwidth]{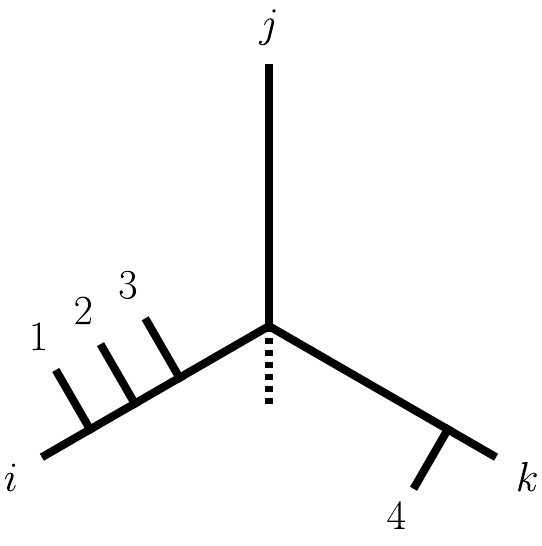} &
    \includegraphics[width=0.3\textwidth]{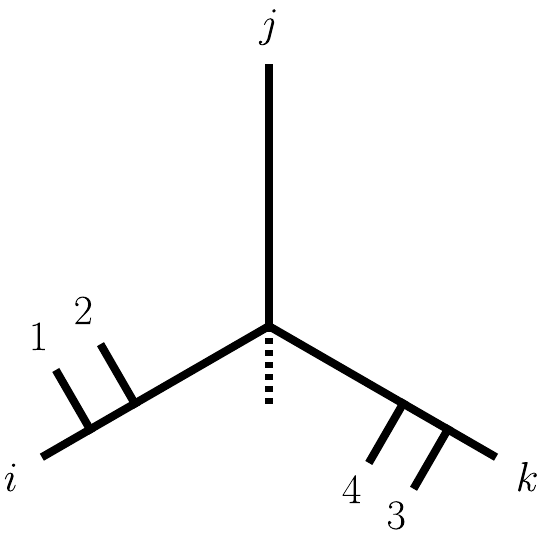}
    \\[-4mm] (g) & (h) & (i)
    \\[2mm]
    \end{tabular}
    \caption{Triskelia with four denominators: (a) $\triskelion400$
    (b) $\triskelion310$ (c) $\triskelion220$ (d) $\triskelion211$
    (e) $\triskelion130$ (f) $\triskelion121$ (g) $\triskelion040$
    (h) $\triskelion031$ (i) $\triskelion022$. Solid external
    lines are either massless or massive, while internal lines are
    all massless.  The dotted line represents either no external
    leg, or either a massless or massive external leg.}
    \label{fig:fourDenominatorTriskelia}
\end{figure}

\section{Survey of Numerator-Free Real-Emission Integrand Reductions}
\label{ResultsSection}

\begin{table}[th]
    \centering
    \renewcommand{\arraystretch}{1.35}
    \begin{tabular}{|c|c|c|c|c|}
\hline
    ~Triskelion~ & ~$r$-indep.~ & ~mass-indep.~ & 
     ~max.{} $\lamh$ power~ & ~max.{} $\tau$ power~ \\
     \hline 
 $\triskelion500$ & $\checkmark$ & $\checkmark$ & 0 & 0\\ \hline
 $\triskelion050$ & $\checkmark$ & $\checkmark$ & 0 & 0\\ \hline
 $\triskelion410$ & $\checkmark$ & $\checkmark$ & 1 & 2\\ \hline
 $\triskelion140$ & $\checkmark$ & $\checkmark$ & 2 & 3\\ \hline
 $\triskelion041$ & $\checkmark$ & $\checkmark$ & 2 & 3\\ \hline
 $\triskelion320$ & $\checkmark$ & $\checkmark$ & 2 & 3\\ \hline
 $\triskelion230$ & $\checkmark$ & $\checkmark$ & 3 & 3\\ \hline
 $\triskelion032$ & $\checkmark$ & $\checkmark$ & 3 & 4\\ \hline
 $\triskelion311$ & $\checkmark$ & $\checkmark$ & 2 & 3\\ \hline
 $\triskelion131$ & $\checkmark$ & $\checkmark$ & 3 & 4\\ \hline
 $\triskelion221$ & $\checkmark$ & $\checkmark$ & 3 & 4\\ \hline
 $\triskelion122$ & $\checkmark$ & $\checkmark$ & 3 & 4\\ \hline
    \end{tabular}
    \caption{Reduction of terms with five-factor denominators,
    for all distinct triskelia.  The second column indicates whether
    all coefficients are independent of $k_r$; the third column whether
    that is independent of the masses of the fixed external legs and
    the momentum flowing into the central vertex.  The fourth and fifth
    columns show the maximum powers of $\lamh$ and $\tau$ in the
    partial-fraction coefficients.}
    \label{NumeratorFreeTable}
\end{table}

In \sect{RealEmissionNumeratorFreeReductionSection}, we gave an
example of a five-factor denominator where we
needed the machinery of computational algebraic geometry in
to demonstrate its reducibility to a sum of four-denominator terms.
We classified the distinct types of denominators we must
consider for general tree amplitudes for any number of
external legs in the previous section, using the notion
of triskelia.  Here, we summarize the results for all the 
different triskelia that arise.  

It is sufficient to examine
the configurations for five-factor denominators.   When examining
terms in amplitudes with more than five factors in denominators ---
that is, when examining real-emission amplitudes with more than 
eight external legs --- we can always choose a subset of the
denominator factors to reduce first.  In general, configurations
with four or fewer denominators cannot necessarily be
partial-fractioned in a $D$-dimensional way.

Each reduction step will
eliminate factors in the denominator, effectively yielding
a term in a lower-point amplitude but with external legs massive
rather than massless.  We must therefore consider all possible
configurations of massive or massless external legs, and
also different configurations of external legs attached to
the `central' vertex: no leg, a massless leg, or a massive leg. 
In \fig{fig:fiveDenominatorTriskelia}, we display graphically
the set of five-denominator skeleton triskelia. 
As it turns out, the key aspects of the algebraic geometry are
independent of these details.  The results of our case-by-case
analysis are shown in \tab{NumeratorFreeTable}.  For all
triskelia, we obtain the desired reduction to a sum of
four-factor denominators, independent of the masses of the
external legs.  The resulting numerators are free of $k_r$
in the form of $s_{xr}$, but 
do depend implicitly on $k_r$ through $\lamh$ and $\tau$.
The latter dependence can be thought of as merely changing
the functional form of the Jacobian~\eqref{InverseAntennaJacobian}.
The maximum powers of $\lamh$ and $\tau$ in the coefficients
of the resulting four-denominator terms are also shown in
the table.

\section{Reduction of Integrand Numerators}
\label{NumeratorReductionSection}

In~\sects{LoopNumeratorFreeReductionSection}{LoopCAGReductionSection},
we examined the reduction of numerator-free loop integrands
in a conventional approach and also as recast in the
language of computational algebraic geometry.  Of course,
in seeking reductions to a basis for virtual contributions,
we must also consider loop integrands with numerators dependent
on the loop momentum.  In general, the loop momentum will be
contracted into either external momenta, external polarization
vectors, or other four-dimensional vectors.  For any
given integral with five or more external legs, 
we can express vectors that aren't external
momenta in terms of a standard basis of four external momenta.
The numerator is then a polynomial in invariants of the loop
momentum contracted with external momenta.  Its degree is
determined by the power counting of the theory; for Yang--Mills theory,
we must consider polynomials of up to degree $n$ for
an $n$-point amplitude.

The conventional approach observes that we can write any of
the invariants,
\begin{equation}
    \ell\cdot k_i = \onehalf \bigl[(\ell-K-k_i)^2-(\ell-K)^2
                                   -(K-k_i)^2+K^2\bigr]\,.
\end{equation}
The first two terms cancel denominators, giving rise to lower-point
integrals, while the last two terms yield integrals with 
a lower-degree numerator.  Repeating this partial-fractioning
ultimately reduces any integral either to a numerator-free
integral, or to a four- (or lower-) point integral.

In order to rephrase this argument in the language of 
algebraic geometry, consider the integrand of a five-point
one-loop integral with external momenta in four dimensions.
The denominators again take the form given in
\eqn{DenominatorForm}, but there are only five instead of six of them.

\def\Poly{\mathop{\rm Poly}\nolimits}
We consider an integrand of the form,
\begin{equation}
    \frac{\Poly(\ell\cdot k_i)}{D_1 D_2 D_3 D_4 D_5}\,,
\end{equation}
where $\Poly$ is a polynomial.  We take as variables
the loop momentum contracted with four specified external
momenta $\{k_{1,2,3,4}\}$, 
and re-express the polynomial in terms of those
variables.  We label this set of variables $\varset{\ell:4}$.
Invariants of external momenta are parameters, that
is constants with respect to the variables.  The statement of
reduction is then that the remainder of the polynomial with
respect to the Gr\"obner basis of the denominators is 
constant.  For this to be true for any polynomial, it is necessary
and sufficient for,
\begin{equation}
    v\mod \GB{}(\{D_i\}_{i=1}^5;\varset{\ell:4}) = \textrm{constant\/}
    \quad\forall v\in \varset{\ell:4}\,.
    \label{LoopReducibility}
\end{equation}
For this to hold, here the Gr\"obner basis must of course
be nontrivial, and the underlying ideal must be zero-dimensional.
Zero dimensionality implies a finite number of solutions;
here there is a unique solution, so
that no Lorentz invariants of the loop momentum can appear on
the right-hand side.  In the loop case, there is
a unique solution in appropriate variables, but this is not
essential to the reduction.

\def\omegan{\omega^{N}}
\Eqn{LoopReducibility} allows us to write a decomposition of
the form,
\begin{equation}
    \frac{\Poly_1(\ell\cdot k_i)}{D_1 D_2 D_3 D_4 D_5} =
   \sum_{j=1}^5\frac{\omegan_j}{D_1\cdots \Xed{D_j}\cdots D_5} 
+ \frac{\omegan_0}{D_1 D_2 D_3 D_4 D_5}\,,
\end{equation}
with $\omegan_0$ independent of $\ell$,
when $\Poly_1$ is linear in $\ell$.  In the loop case,
the coefficients $\omegan_j$ also turn out to be independent of
$\ell$.

\def\omeganr{\omega^{N,r}}
What about the real-emission case?  Instead of the five denominators
we considered in \sect{RealEmissionNumeratorFreeReductionSection},
we now consider terms with just four denominators $\{T_i\}_{i=1}^4$,
using the same set of variables $V$ given in
\eqn{RealEmissionVariableSet}.
We
require the analog of \eqn{LoopReducibility}, namely that,
\begin{equation}
    v\mod \GB{}(\{T_1,T_2,T_3,T_4,R_\tau,R_{\lamh}\};V) = \textrm{Poly}(\tau,\lamh)
    \quad\forall v\in \{s_{\ah r}, s_{r\bh}, s_{r1}, s_{r2}\}\,.
    \label{RealEmissionReducibility}
\end{equation}
We must again require that the Gr\"obner basis be nontrivial,
and that the underlying ideal be zero-dimensional.
We again check this equation for all different configurations; here
that is given by the triskelia with four denominators.  
These are shown in \fig{fig:fourDenominatorTriskelia}. We
find that this condition is obeyed by all, as shown in
\Tab{NumeratorResultTable}.  This allows us to write
the decomposition,
\begin{equation}
        \frac{\Poly_1(r\cdot k_i)}{T_1 T_2 T_3 T_4} =
   \sum_{j=1}^4\frac{\omeganr_j}{T_1\cdots \Xed{T_j}\cdots T_4} 
+ \frac{\omeganr_0}{T_1 T_2 T_3 T_4}\,,
\label{PartialFractionFourDenominators}
\end{equation}
where $\omeganr_0$ is a polynomial in $\tau$ and $\lamh$, with
maximum powers given in \Tab{NumeratorResultTable} for the
different triskelia.  Unlike the loop case, however, we cannot
always ensure that the $\omeganr_j$ are independent of $k_r$
(even after reducing the coefficients with respect to the
Gr\"obner basis of their syzygies, following the same
discussion as below \eqn{SyzygyShift}).  The $r$-independence or
dependence of the coefficients is shown in the last column
of \Tab{NumeratorResultTable}.  
In \eqn{PartialFractionFourDenominators}, we have dropped
terms proportional to $\Rtau$ and $\Rlamh$ as these functions
will vanish on physical configurations.

\def\cross{\text{\sffamily X}}
\def\omitted{\textbf{\Large--}}
\begin{table}[h]
    \centering
    \renewcommand{\arraystretch}{1.35}
    \begin{tabular}{|c|c|c|c|c|c|}
\hline
    ~Triskelion~ &  ~$r$-indep.~ &
     ~mass-indep.~ & 
     ~max.{} $\lamh$ power~ & ~max.{} $\tau$ power~
     & ~r-indep.{} coefficients~ \\
     \hline 
$\triskelion400$ & $\checkmark$ & $\checkmark$ & 0 & 0 & 
$\checkmark$\\ \hline
$\triskelion310$ & $\checkmark$ & $\checkmark$ & 2 & 2 & 
$\checkmark$\\ \hline
$\triskelion220$ & $\checkmark$ & $\checkmark$ & 3 & 3 & 
$\checkmark$\\ \hline
$\triskelion211$ & $\checkmark$ & $\checkmark$ & 3 & 3 & 
$\checkmark$\\ \hline
$\triskelion121$ & $\checkmark$ & $\checkmark$ & 3 & 3 & 
$\checkmark$\\ \hline
$\triskelion022$ & $\checkmark$ & $\checkmark$ & 3 & 3 & 
$\checkmark$\\ \hline
\hline
$\triskelion130$ & $\checkmark$ & $\checkmark$ & 3 & 3 & 
$\cross$\\ \hline
$\triskelion040$ & $\checkmark$ & $\checkmark$ & 0 & 0 & 
$\cross$\\ \hline
$\triskelion031$ & $\checkmark$ & $\checkmark$ & 2 & 2 & 
$\cross$\\ \hline
    \end{tabular}
    \caption{Reduction of four-denominator terms with $k_r$-dependent 
    numerators, for all distinct
    triskelia.  The second column indicates whether $\omeganr_0$
    is independent of $k_r$; the third column whether that is
    independent of the masses of fixed external momenta and the
    momentum flowing into the central vertex.  The 
    fourth and fifth columns give the maximal powers of $\lamh$ 
    and $\tau$ in $\omeganr_0$.  The sixth column indicates whether
    the coefficients $\omeganr_j$ are independent of $k_r$.}
    \label{NumeratorResultTable}
\end{table}

\section{Discussion and Conclusions}
\label{ConclusionSection}

The modern machinery of loop calculations relies on the
existence of a basis of master integrals for each process.
At one loop, this basis is well understood, and the integrals
that arise are process-independent.  The existence of the basis
both at one loop and at higher loops depends in part on
partial-fractioning identities for integrands of general
Feynman integrals.  

In this article, we have taken the first step in constructing
an analogous basis for the real-emission contributions.  We
studied the tree amplitudes out of which these contributions
are computed at next-to-leading order.  We showed the existence
of partial-fractioning identities reducing any term in any
tree-level Feynman diagram to denominators with four or
fewer factors.  Furthermore, each term will have
either a numerator free of explicit dependence on the
emitted particle, or a denominator with three or fewer factors.
We also showed how to construct the required partial-fractioning
identities explicitly.

Our construction relies on the inverse of the antenna mapping,
expressing a color-ordered set of three momenta sufficient to
describe all soft and collinear singularities in a slice of
phase space, in terms of physical proto-jets and an emitted
momentum.  We employ techniques from computational algebraic
geometry in order to find the partial-fractioning identities.

The work described here leaves to future work
the question of delineating 
the form of the basis for the complete integrand
of the real-emission contribution.  This contribution
is given by the square of the tree amplitude.  In addition,
the form of the master-integral basis at one loop also
depends on Lorentz-invariance identities, related to the
vanishing of integrals of certain nontrivial integrands.
The form of the basis at higher loops relies on the vanishing
of a broader class of integrals, generated by total derivatives.
We leave the extension of these ideas to phase-space integrals
to future work as well.

\acknowledgments
We thank Gr\'egory Soyez for help with references.
DAK's work was supported in part by the French 
\textit{Agence 
Nationale pour la Recherche\/}, under grant ANR--17--CE31--0001--01.
He also acknowledges support by the European Research Council, under
grant ERC--AdG--885414. 
BP's work was supported in part by the mentioned ANR grant, 
in part by
the mentioned ERC grant, and in part by the 
the European Union’s Horizon 2020 research and innovation program under the Marie Skłodowska-Curie grant agreement No.~896690 (LoopAnsatz).  
DAK thanks the Kavli Institute for Theoretical Physics, Santa Barbara,
where part of this work was carried out, for its hospitality,
and acknowledges the support in part by the National Science Foundation under Grant No.~NSF PHY--1748958.

\appendix

\section{Jacobian for Change to Protojet Variables}
\label{JacobianAppendix}

\def\xh{{\hat x}}
\def\yh{{\hat y}}
\def\zh{{\hat z}}
There are two factors in the change of variables from the 
partonic momenta
$\{k_i,k_j,k_k\}$ to the protojet momenta $\{k_{\ah},k_r,k_{\bh}\}$:
that contributed 
directly by the measure
$d^D k_i$, and that contributed by the on-shell delta functions.  
Because the singular
momentum does not change form ($k_j=k_r$), 
its Jacobian is 1, and we can focus 
on the remaining variables.
Writing the partonic variables in the form,
\begin{equation}
k_x = \tc_{x,\xh} k_\xh + \tc_{x,r} k_r\,,
\end{equation}
where $x\in\{i,k\}$ and $\xh=\ah,\bh$, the partial derivatives themselves split up into two terms,
\begin{equation}
\begin{aligned}
\frac{\partial k_x^\mu}{\partial k_{\xh}^\nu} &= \tc_{x,\xh} \eta^{\mu}{}_{\nu} 
+ \biggl(\frac{\partial \tc_{x,\yh}}{\partial s_{\zh r}} k_\yh^\mu 
          +\frac{\partial \tc_{x,r}}{\partial s_{\zh r}} k_r^\mu\biggr)\,\frac{\partial s_{\zh r}}{\partial k_{\xh}^\nu}
\\ &= \tc_{x,\xh} \eta^{\mu}{}_{\nu} 
+ 2\biggl(\frac{\partial \tc_{x,\yh}}{\partial s_{\xh r}} k_\yh^\mu 
          +\frac{\partial \tc_{x,r}}{\partial s_{\xh r}} k_r^\mu\biggr)\,k_{r\,\nu}\,.
\end{aligned}
\end{equation}
The form of the second term relies on the fact that 
$\tc_{x,\xh}$ and $\tc_{x,r}$ depend
only on $K^2$, $s_{\ah r}$, and $s_{r\bh}$, with $K^2$ taken to 
be constant.  The form of the
second term means that it is a rank-1 matrix, so that the 
determinant is linear in it.
Define,
\begin{equation}
\begin{aligned}
A_{IJ} &= \tc_{x,\xh} \,\eta^{\mu}{}_{\nu}\,,
\\ B_{IJ} &=
2\biggl(\frac{\partial \tc_{x,\yh}}{\partial s_{\xh r}} k_\yh^\mu 
          +\frac{\partial \tc_{x r}}{\partial s_{\xh r}} k_r^\mu\biggr) 
          \,k_{r\,\nu}\,,
\end{aligned}
\label{TwoMatrices}
\end{equation}
with $I$ denoting the $x$ and $\mu$ indices (taken tensorially), and 
$J$ the $\xh$ and $\nu$ ones.  The Jacobian we want is given by the
determinant,
\begin{equation}
\begin{aligned}
\det\biggl(\frac{\partial k_x^\mu}{\partial k_{\xh}^\nu}\biggr)
&=\det(A+B)
\\ &= \exp\bigl[\ln \det(A+B)\bigr]
\\ &= \exp\bigl[\Tr\ln(A+B)\bigr]
\\ &= \exp\bigl[\Tr\ln(A)+\Tr\ln(1+A^{-1}B)\bigr]
\\ &= \exp\bigl[\Tr\ln(A)+\Tr A^{-1}B + \cdots\bigr]
\\ &= \det A\,\bigl(1+\Tr A^{-1} B\bigr)\,.
\end{aligned}
\end{equation}
The inverse of $A$ is,
\begin{equation}
(A^{-1})_{IJ} = (\tc^{-1})_{x,\xh}\,\eta^{\mu}{}_{\nu}\,.
\end{equation}
The contraction with $\eta$ implies that the second term in the 
expression for $B$
in \eqn{TwoMatrices} disappears.  Our desired determinant is then,
\begin{equation}
\begin{aligned}
\det&\left(\begin{matrix}\tc_{i,\ah}&\tc_{i,\bh}\\
                  \tc_{k,\ah}&\tc_{k,\bh}\end{matrix}\right)^D
\left\{1+\Tr\left[\left(\begin{matrix}\tc_{i,\ah}&\tc_{i,\bh}\\
 \tc_{k,\ah}&\tc_{k,\bh}\end{matrix}\right)^{-1}\left(\begin{matrix}
 \frac{\partial \tc_{i,\yh}}{\partial s_{\ah r}} s_{\yh r}
  &\frac{\partial \tc_{i,\yh}}{\partial s_{r\bh}} s_{\yh r}
  \\ \frac{\partial \tc_{k,\yh}}{\partial s_{\ah r}} s_{\yh r}
  &\frac{\partial \tc_{k,\yh}}{\partial s_{r\bh}} s_{\yh r}
 \end{matrix}\right)\right]\right\} =
\\ &w_0^D(s_{\ah r},s_{r\bh})\tau^D(s_{\ah r},s_{r\bh})
\\ &\hspace*{10mm}\times\left\{1+w_0^{-1}(s_{\ah r},s_{r\bh})
  \tau^{-1}(s_{\ah r},s_{r\bh}) s_{\yh r}
\Tr\left[\left(\begin{matrix}\tc_{k,\bh}\!&\!-\tc_{i,\bh}\\
  -\tc_{k,\ah}\!&\!\tc_{i,\ah}\end{matrix}\right)\!\left(\begin{matrix}
      \frac{\partial \tc_{i,\yh}}{\partial s_{\ah r}}
      &\frac{\partial \tc_{i,\yh}}{\partial s_{r \bh}}
  \\ \frac{\partial \tc_{k,\yh}}{\partial s_{\ah r}}
      &\frac{\partial \tc_{k,\yh}}{\partial s_{r\bh}}
                  \end{matrix}\right)\right]\right\} 
\\ &= w_0^D(s_{\ah r},s_{r\bh})\tau^D(s_{\ah r},s_{r\bh})
\biggl\{1+w_0^{-1}(s_{\ah r},s_{r\bh})
\tau^{-1}(s_{\ah r},s_{r\bh}) s_{\yh r}
\\&\hspace*{45mm}\times
\biggl(
  \tc_{k,\bh}\frac{\partial \tc_{i,\yh}}{\partial s_{\ah r}} 
 -\tc_{i,\bh}\frac{\partial \tc_{k,\yh}}{\partial s_{\ah r}}
 +\tc_{i,\ah}\frac{\partial \tc_{k,\yh}}{\partial s_{r\bh}}
 -\tc_{k,\ah}\frac{\partial \tc_{i,\yh}}{\partial s_{r\bh}} 
 \biggr)\biggr\}\,.
\end{aligned}
\end{equation}
Using the relations arising from momentum conservation~\eqref{MomentumConservationCoefficients},
\begin{equation}
\frac{\partial \tc_{k,\yh}}{\partial s_{\xh r}} = 
-\frac{\partial \tc_{i,\yh}}{\partial s_{\xh r}}\,,
\end{equation}
the first factor in the Jacobian simplifies to,
\begin{equation}
\begin{aligned}
&w_0^D(s_{\ah r},s_{r\bh})\tau^D(s_{\ah r},s_{r\bh})
\biggl\{1+w_0^{-1}(s_{\ah r},s_{r\bh})
\tau^{-1}(s_{\ah r},s_{r\bh}) 
\,s_{\yh r}\biggl[
  \frac{\partial \tc_{i,\yh}}{\partial s_{\ah r}} 
 - \frac{\partial \tc_{i,\yh}}{\partial s_{r\bh}} 
        \biggr]\biggr\}\,.
\end{aligned}
\end{equation}

A few side computations show that,
\begin{equation}
\begin{aligned}
K^2\frac{\partial w_0(s_{\ah r},s_{r\bh})}{\partial s_{\ah r}} &= -1\,,
\\ K^2\frac{\partial w_0(s_{\ah r},s_{r\bh})}{\partial s_{r\bh}} &= -1\,,
\\ 2 K^2\frac{\partial w_\Sigma(s_{\ah r},s_{r\bh})}{\partial s_{\ah r}}
&=  \frac{\partial \lamh(s_{\ah r},s_{r\bh})}{\partial s_{\ah r}}
(s_{\ah r}+s_{r\bh})
   +\lamh(s_{\ah r},s_{r\bh})+1\,,
\\ 2 K^2\frac{\partial w_\Sigma(s_{\ah r},s_{r\bh})}{\partial s_{r\bh}} &= 
   \frac{\partial \lamh(s_{\ah r},s_{r\bh})}{\partial s_{r\bh}}
   (s_{\ah r}+s_{r\bh})
   +\lamh(s_{\ah r},s_{r\bh})-1\,,
\\
\frac{4 (K^2)^2}{\tau^3(s_{\ah r},s_{r\bh})} \frac{\partial \tau(s_{\ah r},s_{r\bh})}{\partial s_{\ah r}} &= 
\\ & \hspace*{-15mm} 2 K^2 - s_{\ah r} + s_{r\bh} - 2(K^2 - s_{\ah r})\lamh(s_{\ah r},s_{r\bh})
\\&\hspace*{-15mm} 
- (s_{\ah r} + s_{r\bh})\lamh{}^2(s_{\ah r},s_{r\bh}) 
  - (s_{\ah r} + s_{r\bh})^2\lamh(s_{\ah r},s_{r\bh})\frac{\partial \lamh(s_{\ah r},s_{r\bh})}{\partial s_{\ah r}}
\\&\hspace*{-15mm} 
-  (2 K^2 - s_{\ah r} - s_{r\bh}) (s_{\ah r} - s_{r\bh})
  \frac{\partial \lamh(s_{\ah r},s_{r\bh})}{\partial s_{\ah r}}
\,,
\\
\frac{4 (K^2)^2}{\tau^3(s_{\ah r},s_{r\bh})} \frac{\partial \tau(s_{\ah r},s_{r\bh})}{\partial s_{r\bh}} &= 
\\ & \hspace*{-15mm} 2 K^2 + s_{\ah r} - s_{r\bh} + 2(K^2 - s_{r\bh})\lamh(s_{\ah r},s_{r\bh})
\\&\hspace*{-15mm} 
- (s_{\ah r} + s_{r\bh})\lamh{}^2(s_{\ah r},s_{r\bh}) 
  - (s_{\ah r} + s_{r\bh})^2\lamh(s_{\ah r},s_{r\bh})\frac{\partial \lamh(s_{\ah r},s_{r\bh})}{\partial s_{r\bh}}
\\&\hspace*{-15mm} 
-  (2 K^2 - s_{\ah r} - s_{r\bh}) (s_{\ah r} - s_{r\bh})
  \frac{\partial \lamh(s_{\ah r},s_{r\bh})}{\partial s_{r\bh}}
\,,
\end{aligned}
\end{equation}
so that,
\begin{equation}
\begin{aligned}
s_{\yh r}\biggl[
  \frac{\partial \tc_{i,\yh}}{\partial s_{\ah r}} 
 - \frac{\partial \tc_{i,\yh}}{\partial s_{r\bh}} 
        \biggr] &= 
        -\tau(s_{\ah r},s_{r\bh})w_0(s_{\ah r},s_{r\bh})
        \\& \hphantom{=!} + \tau^3(s_{\ah r},s_{r\bh}) w_0(s_{\ah r},s_{r\bh})\biggl(
        w_J(s_{\ah r},s_{r\bh})
        \\& \hspace*{48mm}+\frac{s_{\ah r} s_{r\bh}}{K^2}\biggl[\frac{\partial\lamh(s_{\ah r},s_{r\bh})}{\partial s_{\ah r}}
        -\frac{\partial\lamh(s_{\ah r},s_{r\bh})}{\partial s_{r\bh}}\biggr]\biggr)\,.
\end{aligned}
\end{equation}
Putting everything together, we find that,
\begin{equation}
\begin{aligned}
    \frac{d^D k_i}{(2\pi)^4}&\,\frac{d^D k_j}{(2\pi)^4}\,
\frac{d^D k_k}{(2\pi)^4} =
\\&
\tau^{D+2}(s_{\ah r},s_{r\bh}) w_0^{D}(s_{\ah r},s_{r\bh})
\\&\times
\biggl[
 w_J(s_{\ah r},s_{r\bh})
+ \frac{s_{\ah r}s_{r\bh}}{K^2}\Bigl[
 \frac{\partial\lamh(s_{\ah r},s_{r\bh})}{\partial s_{\ah r}}
   -\frac{\partial\lamh(s_{\ah r},s_{r\bh})}{\partial s_{r\bh}}
   \Bigr]\biggr]
\\&\times
\frac{d^D k_\ah}{(2\pi)^4}\,\frac{d^D k_\bh}{(2\pi)^4}\,
\frac{d^D k_r}{(2\pi)^4}\,.
\end{aligned}
\label{MeasureJacobian}
\end{equation}

To determine the Jacobian contribution from the delta functions, we note that (imposing $k_r^2=0$),
\begin{equation}
\begin{aligned}
    k_i^2 &= \tc_{i,\ah}^2 k_\ah^2 + \tc_{i,\bh}^2 k_\bh^2 
       + (K^2-k_\ah^2-k_\bh^2) \tc_{i,\ah} \tc_{i,\bh} + L_{\ah r} \tc_{i,\ah} \tc_{i,r} 
             + L_{r\bh} \tc_{i,\bh} \tc_{i,r}
\\&= \tc_{i,\ah} (\tc_{i,\ah}-\tc_{i,\bh}) k_\ah^2 + \tc_{i,\bh} (\tc_{i,\bh}-\tc_{i,\ah}) k_\bh^2 
       + K^2 \tc_{i,\ah} \tc_{i,\bh} + L_{\ah r} \tc_{i,\ah} \tc_{i,r} 
             + L_{r\bh} \tc_{i,\bh} \tc_{i,r}
\,,
\end{aligned}
\end{equation}
where the $L_{mn} = 2 k_m\cdot k_n$ are treated as independent
(and $K^2$ is constant).
The last three terms on the second line are independent of
$k_{\ah,\bh}^2$.  A similar form holds for $k_k^2$.
The Jacobian factor is then,
\begin{equation}
\det{}^{-1}\left(\begin{matrix}
\tc_{i,\ah} (\tc_{i,\ah}-\tc_{i,\bh}) & \tc_{i,\bh} (\tc_{i,\bh}-\tc_{i,\ah})\\
\tc_{k,\ah} (\tc_{k,\ah}-\tc_{k,\bh}) & \tc_{k,\bh} (\tc_{k,\bh}-\tc_{k,\ah})
\end{matrix}\right)
= \tau^{-3}(s_{\ah r},s_{r\bh}) w_0^{-3}(s_{\ah r},s_{r\bh}) \,.
\end{equation}
That is,
\begin{equation}
\begin{aligned}
  \deltap(k_i^2)\,\deltap(k_j^2)\,\deltap(k_k^2) &=
\tau^{-3}(s_{\ah r},s_{r\bh}) w_0^{-3}(s_{\ah r},s_{r\bh})
\,  \deltap(k_{\ah}^2)\,\deltap(k_{\bh}^2)\,\deltap(k_r^2)
\,.
\end{aligned}
\label{DeltaFunctionJacobian}
\end{equation}
We have checked the product of \eqns{MeasureJacobian}{DeltaFunctionJacobian} numerically.

\section{Finite-Field Momentum Configuration}
\label{SampleConfigurationAppendix}

For the calculations described 
in \sectst{LoopCAGReductionSection}{ResultsSection}{NumeratorReductionSection}, 
we used the following momentum configuration,
\begin{equation}
    \begin{aligned}
    k_1 &= (213587426, 225063685, 46141879, 775249)\,,\\
    k_2 &= (249027085, 4905933, 178795165, 85542514)\,,\\
    k_3 &= (260690655, 249277361, 167348189, 63632136)\,,\\
    k_4 &= (200153951, 144210099, 200969668, 71069250)\,,\\
    k_5 &= (190546565, 103074265, 231179370, 117909270\,,\\
    k_{\ah} &= (143139312, 121865425, 85376447, 146859458)\,,\\
    k_{\bh} &= (120877711, 254021396, 192607446, 65421205)\,,\\
    \end{aligned}
\end{equation}
in the finite field $\mathbb{Z}_p$, with $p=275604541$.

The symbolic coefficients $c_i$ are then,
\begin{equation}
\begin{array}{rlrlrlrl}
n_1 &= 68813401\,,&\quad
n_2 &= 66817369\,,&\quad
n_3 &= 143203095\,,&\quad
n_4 &= 163372679\,,\\
n_5 &= 129247152\,,&\quad
n_6 &= 104659326\,,&\quad
n_7 &= 170945215\,,&\quad
n_8 &= 73713437\,,\\
n_9 &= 191284565\,,&\quad
n_{10} &= 59757494\,,&\quad
n_{11} &= 107723833\,,&\quad
n_{12} &= 234166597\,,\\
n_{13} &= 63221382\,,&\quad
n_{14} &= 63221382\,,&\quad
n_{15} &= 65226064\,,&\quad
n_{16} &= 65226064\,,\\
n_{17} &= 156588253\,,&\quad
n_{18} &= 119016288\,,&\quad
n_{19} &= 119016288\,,&\quad
n_{20} &= 119016288\,,\\
n_{21} &= 43709843\,,&\quad
n_{22} &= 43709843\,,&\quad
n_{23} &= 232255515\,,&\quad
n_{24} &= 43349026\,,\\
n_{25} &= 43349026\,,&\quad
n_{26} &= 43349026\,,&\quad
n_{27} &= 50900458\,,&\quad
n_{28} &= 44778046\,,\\
n_{29} &= 8669401\,,&\quad
n_{30} &= 112897039\,,&\quad
n_{31} &= 111335751\,,&\quad
n_{32} &= 251337127\,,\\
n_{33} &= 24267414\,,&\quad
n_{34} &= 82519587\,,&\quad
n_{35} &= 183168880\,,&\quad
n_{36} &= 185185937\,,\\
n_{37} &= 227237491\,,&\quad
n_{38} &= 96901878\,,&\quad
n_{39} &= 72634464\,,&\quad
n_{40} &= 72634464\,,\\
n_{41} &= 69153044\,,&\quad
n_{42} &= 69153044\,,&\quad
n_{43} &= 262627742\,,&\quad
n_{44} &= 12976799\,,\\
n_{45} &= 12976799\,,&\quad
n_{46} &= 12976799\,,&\quad
n_{47} &= 263569323\,,&\quad
n_{48} &= 263569323\,,\\
n_{49} &= 251033983\,,&\quad
n_{50} &= 24570558\,,&\quad
n_{51} &= 24570558\,,&\quad
n_{52} &= 24570558\,,\\
n_{53} &= 254554542\,,&\quad
n_{54} &= 91321399\,,&\quad
n_{55} &= 17338801\,,&\quad
n_{56} &= 165039173\,,\\
n_{57} &= 138306088\,,&\quad
n_{58} &= 251534105\,,&\quad
n_{59} &= 49964064\,,&\quad
n_{60} &= 11715612\,,\\
n_{61} &= 109876218\,,&\quad
n_{62} &= 17200224\,,&\quad
n_{63} &= 102454141\,,&\quad
n_{64} &= 59908817\,,\\
n_{65} &= 130084877\,,&\quad
n_{66} &= 10351656\,,&\quad
n_{67} &= 20144505\,,&\quad
n_{68} &= 102106340\,,\\
n_{69} &= 10351656\,,&\quad
n_{70} &= 100838991\,,&\quad
n_{71} &= 135520768\,,&\quad
n_{72} &= 124166887\,,\\
n_{73} &= 43415258\,,&\quad
n_{74} &= 23593835\,,&\quad
n_{75} &= 85776851\,,&\quad
n_{76} &= 136861535\,,\\
n_{77} &= 17321109\,,&\quad
n_{78} &= 63204303\,,&\quad
n_{79} &= 16284639\,,&\quad
n_{80} &= 53293007\,,\\
n_{81} &= 92040307\,,&\quad
n_{82} &= 135526622\,,&\quad
n_{83} &= 91586436\,,&\quad
n_{84} &= 92040307\,,\\
n_{85} &= 102472919\,,&\quad
n_{86} &= 24177935\,,&\quad
n_{87} &= 68593640\,,&\quad
n_{88} &= 61599662\,,\\
n_{89} &= 36018113\,,&\quad
n_{90} &= 24365765\,,&\quad
n_{91} &= 53307569\,,&\quad
n_{92} &= 85606915\,,\\
n_{93} &= 104341\,,&\quad
n_{94} &= 53307569\,,&\quad
n_{95} &= 45334800\,,&\quad
n_{96} &= 42833228\,,\\
n_{97} &= 81803435\,,&\quad
n_{98} &= 4829961\,,&\quad
n_{99} &= 135519863\,.
\end{array}
\end{equation}


\end{document}